\documentclass[10pt,twocolumn,preprintnumbers,amsmath,amssymb,nofootinbib
,superscriptaddress]{revtex4-1}
\usepackage{graphicx,longtable,mathrsfs,color,array}
\usepackage[hidelinks]{hyperref}
\usepackage[usenames,dvipsnames]{xcolor} 
\usepackage{amssymb,amsmath,mathtools,mathrsfs,slashed} 
\usepackage{epsfig,subfigure,placeins,float} 
\usepackage{booktabs,longtable,ctable,multirow} 
\usepackage{exscale,relsize} 
\usepackage[normalem]{ulem} 
\usepackage{enumerate}
\usepackage{times, mathptmx} 
\usepackage[utf8]{inputenc}
\usepackage{color}
\usepackage{hyperref}
\usepackage{graphicx}
\usepackage{color}
\usepackage[mathscr]{euscript}
\usepackage{graphicx,graphics}
\usepackage{bm}
\usepackage{tabularx}
\allowdisplaybreaks[1]

\begin{document}

\title{Quasi-Normal Modes of Hairy Scalar Tensor Black Holes: Odd Parity}
\author{Oliver J. Tattersall}
\email{oliver.tattersall@physics.ox.ac.uk}
\affiliation{Astrophysics, University of Oxford, DWB, Keble Road, Oxford OX1 3RH, UK}

\date{Received \today; published -- 00, 0000}

\begin{abstract}
The odd parity gravitational Quasi-Normal Mode spectrum of black holes with non-trivial scalar hair in Horndeski gravity is investigated. We study `almost' Schwarzschild black holes such that any modifications to the spacetime geometry (including the scalar field profile) are treated as small quantities. A modified Regge-Wheeler style equation for the odd parity gravitational degree of freedom is presented to quadratic order in the scalar hair and spacetime modifications, and a parameterisation of the modified Quasi-Normal Mode spectrum is calculated. In addition, statistical error estimates for the new hairy parameters of the black hole and scalar field are given.
\end{abstract}

\maketitle
\section{Introduction}

Gravitational wave (GW) astronomy is now in full swing, thanks to numerous and frequent observations of compact object mergers by advanced LIGO and VIRGO \cite{LIGOScientific:2018mvr}. With next generation ground and space based GW detectors on the horizon, the prospect of performing black hole spectroscopy (BHS) \cite{Dreyer:2003bv,Berti:2005ys,Gossan:2011ha,Meidam:2014jpa,Berti:2016lat,Berti:2018vdi,Baibhav:2018rfk,Giesler:2019uxc,Bhagwat:2019dtm,Bhagwat:2019bwv,Maselli:2019mjd,Ota:2019bzl,Cabero:2019zyt} (the gravitational analog to atomic spectroscopy) is tantalisingly close. With BHS, one aims to discern multiple distinct frequencies of gravitational waves emitted during the ringdown of the highly perturbed remnant black hole of a merger event. 

These frequencies, known as Quasi-Normal Modes (QNMs), act as fingerprints for a black hole, being dependent on both the background properties of a black hole (e.g. its mass) and on the laws of gravity \cite{1975RSPSA.343..289C,0264-9381-16-12-201,Kokkotas:1999bd,Berti:2009kk,Konoplya:2011qq}. In General Relativity (GR), the QNM spectrum of a Kerr black hole is entirely determined by its mass and angular momentum, and the black hole is said to have no further `hairs' \cite{Kerr:1963ud,Israel:1967wq,Israel:1967za,Carter:1971zc,1972CMaPh..25..152H,PhysRevD.5.2403}. Thus the detection of \textit{multiple} QNMs in the ringdown portion of a gravitational wave signal allows a consistency check between the inferred values of $M$ and $J$ from each frequency. In \cite{2016PhRvL.116v1101A} the least damped QNM (assumed to be the $\ell=m=2$ fundamental overtone) of the first gravitational wave detection GW150914 is observed. The values of the oscillation frequency and damping time of this mode are consistent with the predictions of GR, assuming a Kerr black hole with the inferred mass and spin of the GW150914 remnant. As mentioned, however, multiple modes must be observed to test the consistency of the GR spectrum.

In gravity theories other than GR, however, the situation can be markedly different. For example, black holes may not be described by the Kerr solution, and may have properties other than mass or angular momentum that affect its QNM spectrum. Such black holes are said to have `hair' and, despite no-hair theorems existing for various facets of modified gravity, finding and studying hairy black hole solutions is at the forefront of strong gravity research \cite{Blazquez-Salcedo:2016enn,Blazquez-Salcedo:2017txk,Silva:2017uqg,Antoniou:2017acq,Antoniou:2017hxj,Bakopoulos:2018nui,Minamitsuji:2018xde,Sullivan:2019vyi,Macedo:2019sem,Konoplya:2001ji,Dong:2017toi,Cardoso:2018ptl,Brito:2018hjh,Franciolini:2018uyq}. On the other hand, even if black holes in modified gravity theories are described by the same background solution as in GR (i.e. they have no hair), their perturbations may obey modified equations of motion that alter the emitted gravitational wave signal \cite{Barausse:2008xv,Molina:2010fb,Tattersall:2017erk,Tattersall:2018nve,Tattersall:2019pvx}. 

In this paper we will investigate the first possibility, where modified gravity black holes are altered from their usual description in GR due to their interactions with new gravitational fields. We will, however, assume that black holes are (to first order at least) well described by the GR solutions, and any modifications to the background spacetime are treated as small quantities. As various observations appear to suggest that black holes are well described by the suite of GR solutions \cite{2016PhRvL.116v1101A,Isi:2019aib}, this approach seems sensible. In this way we can treat the new modified QNM spectrum of these hairy black holes as a small correction to the original GR spectrum, greatly simplifying the analytical and numerical analysis. This is analogous to the study of slowly rotating black holes, where the Kerr background solution is treated as a small modifiation to the Schwarzschild metric (with the dimensionless black hole spin considered as an `expansion' parameter), and gravitational wave perturbations are studied on top of this new background (see, for example, \cite{Kojima:1992ie,Pani:2012zz,Pani:2012bp,Tattersall:2018axd}). In this way, even without full knowledge of an exact black hole solution, we can probe how the QNM spectrum will be affected by modifications to the geometry and scalar profile.

We will specifically focus on the Horndeski family of scalar-tensor theories of gravity \cite{Horndeski:1974wa}, where a new gravitational scalar field interacts non-minimally with the metric. The motivation for working with Horndeski gravity is that it encompasses a large family of scalar-tensor theories of gravity (perhaps the simplest extension to GR that one could imagine), ranging from models which describe dark energy to those inspired by string theory \cite{Kobayashi:2019hrl}. Furthermore, for simplicity, we will restrict ourselves to looking only at the odd parity sector of perturbations to spherically symmetric black holes, i.e. we will assume that the black holes studied here are described by a slightly modified Schwarzschild metric. The extension of this work to the even parity sector of spherically symmetric black holes, and to include the effects of rotation, are left as future exercises.

\textit{Summary}: In section \ref{horndeskisection} we will introduce the action for Horndeski gravity, the hairy black hole metric and scalar field profile that we are considering, and explore the odd parity gravitational perturbations of this system. In section \ref{QNMsection} we will utilise the results of \cite{Cardoso:2019mqo} to calculate the modified QNM spectrum of the modified black hole, and provide observational error estimates for the new hairy parameters given a power law ansatz for the hairy modifications. We will then conclude with a discussion of the results presented here.

Throughout we will use natural units with $G=c=1$, except where otherwise stated. The metric signature will be mostly positive.

\section{Horndeski Gravity}\label{horndeskisection}

\subsection{Background}

A general action for scalar-tensor gravity is given by the Horndeski action \cite{Horndeski:1974wa,Kobayashi:2011nu}:
\begin{align}
S=\int d^4x\sqrt{-g}\sum_{n=2}^5L_n,\label{Shorndeski}
\end{align}
where the component Horndeski Lagrangians are given by:
\begin{align}
L_2&=G_2(\phi,X)\nonumber\\
L_3&=-G_3(\phi,X)\Box \phi\nonumber\\
L_4&=G_4(\phi,X)R+G_{4X}(\phi,X)((\Box\phi)^2-\phi^{\alpha\beta}\phi_{\alpha\beta} )\nonumber\\
L_5&=G_5(\phi,X)G_{\alpha\beta}\phi^{\alpha\beta}-\frac{1}{6}G_{5X}(\phi,X)((\Box\phi)^3 \nonumber\\
& -3\phi^{\alpha\beta}\phi_{\alpha\beta}\Box\phi +2 \phi_{\alpha\beta}\phi^{\alpha\sigma}\phi^{\beta}_{\sigma}),
\end{align}
where $\phi$ is the scalar field with kinetic term $X=-\phi_\alpha\phi^\alpha/2$, $\phi_\alpha=\nabla_\alpha\phi$, $\phi_{\alpha\beta}=\nabla_\alpha\nabla_\beta\phi$, and $G_{\alpha\beta}=R_{\alpha\beta}-\frac{1}{2}R\,g_{\alpha\beta}$ is the Einstein tensor. The $G_i$ are arbitrary functions of $\phi$ and $X$, with derivatives $G_{iX}$ with respect to $X$. GR is given by the choice $G_4=M_{P}^2/2$ with all other $G_i$ vanishing and $M_{P}$ being the reduced Planck mass. 

The Horndeski action is formulated in such a way as to ensure 2$^{nd}$ order-derivative equations of motion, free from any Ostrogradski instability related to higher order time derivatives \cite{Langlois:2015cwa}. Note that eq.~(\ref{Shorndeski}) is \textit{not} the most general action for scalar-tensor theories, and it has been shown that it can be extended to an arbitrary number of terms \cite{Zumalacarregui:2013pma,Gleyzes:2014qga,Gleyzes:2014dya,Achour:2016rkg}.

Horndeski theories can be consistent with solar system tests of gravity through screening mechanisms \cite{Kase:2013uja}, and the conditions for Horndeski theories to have appropriate Einstein gravity limits have been studied in \cite{McManus:2016kxu}. Recently, strong constraints were placed on the Horndeski action by the observation of gravitational waves from a binary neutron star merger GW170817 by LIGO and VIRGO \cite{PhysRevLett.119.161101}, along with its optical counterpart the gamma ray burst GRB 170817A \cite{2041-8205-848-2-L12,2041-8205-848-2-L13,2041-8205-848-2-L14,2041-8205-848-2-L15,2017Sci...358.1556C}. Due to the almost coincident arrival of both gravitational waves and photons at Earth from this event, the speed of gravitational waves was constrained to be within 1 part in $10^{15}$ of that of light, leading to tight constraints on the Horndeski action (as well as other modified gravity theories) \cite{Lombriser:2015sxa,Lombriser:2016yzn,Baker:2017hug,Creminelli:2017sry,Sakstein:2017xjx,Ezquiaga:2017ekz}. Specifically, we require $G_4=G_4(\phi)$ and $G_5=0$ to satisfy the gravitational wave speed constraint.

These constraints were, however, made under the assumption that the Horndeski scalar field plays a \textit{significant} cosmological role. If one is instead content to allow $\phi$ to become unimportant for cosmology (and thus to the propagation of gravitational waves over cosmological distances), we are still free to consider $G_4$ and $G_5$ in their entirety.\footnote{See also, for example, the discussion in \cite{deRham:2018red}, where it is argued that one should be careful when applying gravitational wave constraints to dark energy effective field theories.} This is what we will do in this paper, allowing the full suite of Horndeski functions to be relevant to black hole solutions, whilst knowing that those terms which contribute to the speed excess of gravitational waves must vanish in a cosmological setting. For a discussion of black hole solutions where we \textit{do} require any non-GR fields to maintain a cosmological significance, see \cite{PhysRevD.97.084005}.

For a spherically symmetric black hole solution in Horndeski gravity we assume the following form for the metric $g$ and scalar field $\phi$ in `Schwarzschild-like' coordinates:
\begin{subequations}
\begin{align}
ds^2=&\;g_{\mu\nu}dx^\mu dx^\nu=\;-A(r)dt^2+B(r)^{-1}dr^2+C(r)d\Omega^2\\
\phi=&\;\phi(r)
\end{align}
\end{subequations}
where $d\Omega^2$ is the metric on the unit 2-sphere. 

Our starting point will be a hairless Schwarzschild solution, as in GR, such that $A=B=1-2M/r$, $C=r^2$ and $\phi=\phi_0=const$, where $M$ is the mass of the black hole. We will now introduce small deviations as `hair' in both the spacetime geometry and in the scalar field profile, leading to a modified `almost' Schwarzschild black hole. Using $\epsilon$ as a book keeping parameter to track the order of smallness of the hair, we make the following ansatz to second order in $\epsilon$:
\begin{subequations}\label{ansatz}
\begin{align}
A(r)=&\;B(r)=1-\frac{2M}{r}+\epsilon \delta A_1(r) + \epsilon^2 \delta A_2(r)+\mathcal{O}(\epsilon^3)\label{gbackground}\\
C(r)=&\;\left(1+\epsilon \delta C_1(r) + \epsilon^2 \delta C_2(r)\right) r^2 + \mathcal{O}(\epsilon^3)\\
\phi(r)=&\;\phi_0+\epsilon \delta \phi_1(r)+ \epsilon^2 \delta\phi_2(r)+\mathcal{O}(\epsilon^3),\label{phibackground}
\end{align}
\end{subequations}
where we are remaining agnostic as to the exact form of the modifications, merely supposing that such perturbations could exist.

Note that the functions $\delta A_i$ will shift the location of the horizon from $r=2M$. We will look into this further in a later section. Asymptotically we might expect any $\mathcal{O}(\epsilon)$ terms to decay as $r\to\infty$ so as to recover flat space far away from the black hole. It is also conceivable, however, that modified gravity effects might manifest themselves as an apparent cosmological constant term, leading to an asymptotically de Sitter (or anti-de Sitter) form for the metric.

\subsection{Black Hole Perturbations}\label{perturbsec}

We now consider odd parity perturbations to the `almost Schwarzschild' black hole described by eq.~(\ref{gbackground})~-~(\ref{phibackground}). For simplicity we will only be considering odd parity perturbations, and as such we do not need to consider the coupling of the even parity metric perturbation to the scalar degree of freedom (nor, indeed, perturbations to the effective energy momentum tensor of the scalar field). In this way we can see the effect on the QNM spectrum of the black hole due entirely to the hairy nature of the background, and not to (for example) couplings to scalar field perturbations. An analysis of the even parity sector for perturbatively hairy black holes in Horndeski gravity is left as a future extension to this work; the stability of generic spherically symmetric black holes in Horndeski gravity was studied in \cite{Kobayashi:2012kh,Kobayashi:2014wsa}, whilst \cite{Franciolini:2018uyq} builds an effective field theory for QNMs in scalar-tensor gravity in the unitary gauge.

In the Regge-Wheeler gauge \cite{Regge:1957td}, odd parity perturbations $h_{\mu\nu}$ to the metric $g_{\mu\nu}$ can be decomposed into tensorial spherical harmonics and written in terms of two `perturbation fields' $h_0(r)$ and $h_1(r)$ in the following way:
\begin{widetext}
\begin{align}
 h_{\mu\nu,\ell m}^{\text{odd}}=&
 \begin{pmatrix}
 0&0&-h_0(r)\frac{1}{\sin\theta}\frac{\partial}{\partial\phi}&h_0(r)\sin\theta\frac{\partial}{\partial\theta}\\
 0&0&-h_1(r)\frac{1}{\sin\theta}\frac{\partial}{\partial\phi}&h_1(r)\sin\theta\frac{\partial}{\partial\theta}\\
 -h_0(r)\frac{1}{\sin\theta}\frac{\partial}{\partial\phi}&-h_1(r)\frac{1}{\sin\theta}\frac{\partial}{\partial\phi}&0&0\\
h_0(r)\sin\theta\frac{\partial}{\partial\theta}&h_1(r)\sin\theta\frac{\partial}{\partial\theta}&0&0
 \end{pmatrix}Y^{\ell m}e^{-i\omega t}
\end{align}
\end{widetext}
where $Y^{\ell m}$ is the usual scalar spherical harmonic. Note that in the Regge-Wheeler gauge we have been able to set a third perturbation field $h_2(r)$ (which would have populated the bottom right hand corner of $h_{\mu\nu}^{\text{odd}}$) to zero.

After expanding the action given by eq.~(\ref{Shorndeski}) to second order in the perturbation fields $h_i$, and integrating over $\theta$ and $\phi$, it was shown in \cite{Ganguly:2017ort} that the following action is obtained:
\begin{align}
S^{(2)}=\int dt dr \left[a_1h_0^2+a_2h_1^2+a_3\left(\dot{h}_1^2 h_0^{\prime2}-2\dot{h}_1h_0^\prime+2\frac{C^\prime}{C}\dot{h}_1h_0\right)\right]\label{secondorderaction}
\end{align}
where a dot represents a time derivative and a prime a radial derivative. The coefficients $a_i$ are given in appendix \ref{appendixaction}.

We can see in eq.~(\ref{secondorderaction}) that $h_0$ is an auxiliary field (i.e. without a time derivative). Through further manipulation of eq.~(\ref{secondorderaction}), a redefined field $Q(h_1)$ is shown in \cite{Ganguly:2017ort} to obey the following equation of motion:
\begin{align}
\left[\frac{d^2}{dr_\ast^2}+\frac{\mathcal{F}}{\mathcal{G}}\omega^2-\mathscr{V}\right]Q=0\label{RWgen}
\end{align}
where $r_\ast$ is the tortoise coordinate defined by $dr=\sqrt{AB}dr_\ast$, and the potential $\mathscr{V}$ is given by:
\begin{align}
\mathscr{V}=&\;\ell(\ell+1)\frac{A}{C}\frac{\mathcal{F}}{\mathcal{H}}-\frac{C^2}{4C^\prime}\left(\frac{ABC^{\prime 2}}{C^3}\right)^\prime-\frac{C^2\mathcal{F}^2}{4\mathcal{F}^\prime}\left(\frac{AB\mathcal{F}^{\prime 2}}{C^2\mathcal{F}^3}\right)^\prime \nonumber\\
&-\frac{2A\mathcal{F}}{C\mathcal{H}}.\label{Vgen}
\end{align}
The functions $\mathcal{F}$, $\mathcal{G}$, and $\mathcal{H}$ are combinations of the Horndeski $G_i$ functions evaluated at the level of the background:
\begin{subequations}\label{fgheqs}
\begin{align}
\mathcal{F}=&\;2\left(G_4+\frac{1}{2}B\phi^\prime X^\prime G_{5X}-X G_{5\phi}\right)\\
\mathcal{G}=&\;2\left[G_4-2XG_{4X}+X\left(\frac{A^\prime}{2A}B\phi^\prime G_{5X}+G_{5\phi}\right)\right]\\
\mathcal{H}=&\;2\left[G_4-2XG_{4X}+X\left(\frac{C^\prime}{2C}B\phi^\prime G_{5X}+G_{5\phi}\right)\right].
\end{align}
\end{subequations}
Furthermore note that we have suppressed spherical harmonic indices for compactness, but eq.~(\ref{RWgen}) is assumed to hold for each $\ell$.

Eq.~(\ref{RWgen}) is the analog of the Regge-Wheeler equation \cite{Regge:1957td} for a generic spherically symmetric black hole in Horndeski gravity. Imposing the boundary conditions that gravitational radiation should be purely `ingoing' at the black hole horizon, and purely `outgoing' at spatial infinity, one can find find the discrete spectrum of  QNM frequencies $\omega$ that satisfies eq.~(\ref{RWgen}).

We now Taylor expand all of the terms in eq.~(\ref{RWgen}) to $O(\epsilon^2)$ using eq.~(\ref{gbackground})~-~(\ref{phibackground}) to take into account the effects of the perturbative black hole hair that we introduced in eq.~(\ref{ansatz}), resulting in the following:
\begin{widetext}
\begin{align}
\left[\frac{d^2}{dr_\ast^2}+\omega^2\left(1+\epsilon^2\alpha_{T}(r)\right)-A(r)\left(\frac{\ell(\ell+1)}{r^2}-\frac{6M}{r^3}+\epsilon\delta V_1+\epsilon^2\delta V_2 \right)\right]Q=0\label{RWhairy}
\end{align}
\end{widetext}
where $\alpha_T$ is the speed excess of gravitational waves \cite{DeFelice:2011bh,Bellini:2014fua} given by, to $O(\epsilon^2)$:
\begin{align}
\alpha_T(r)=&\;-\left(1-\frac{2M}{r}\right)\frac{G_{4X}-G_{5\phi}}{G_4}\delta\phi_1^{\prime 2},\label{alphaT}
\end{align}
whilst the potential perturbations are given by:
\begin{widetext}
\begin{subequations}\label{deltaVs}
\begin{align}
\delta V_1=&\;\frac{1}{2r^2}\left[4\delta A_1-2r\delta A_1^\prime-2(\ell+2)(\ell-1)\delta C_1+2(r-3M)\delta C_1^\prime-r(r-2M)\delta C_1^{\prime\prime}-\frac{G_{4\phi}}{G_4}\left(r\left(r-2M\right)\delta\phi_1^{\prime\prime}-2(r-3M)\delta\phi_1^\prime\right)\right]\label{deltaV1}\\
\delta V_2=&\;\frac{1}{4r^2}\left[8\delta A_2-4r\delta A_2^\prime+4(\ell+2)(\ell-1)\left(\delta C_1^2-\delta C_2\right)+3r(r-2M)\delta C_1^{\prime2}+4(r-3M)\delta C_2^\prime-2r(r-2M)\delta C_2^{\prime\prime}+4r\delta A_1\delta C_1^\prime\right.\nonumber\\
&\left.-2r^2\left(\delta A_1^\prime \delta C_1^\prime+\delta A_1\delta C_1^{\prime\prime}\right)-4(r-3M)\delta C_1\delta C_1^{\prime\prime}+2r(r-2M)\delta C_1 \delta C_1^{\prime\prime}\right]\nonumber\\
&-\frac{1}{2r^2}\frac{G_{4\phi}}{G_4}\left[-2(r-3M)\delta\phi_2^\prime + r \left( r\delta A_1^\prime\delta\phi_1^\prime -\delta A_1\left(2\delta \phi_1^\prime - r \delta\phi_1^{\prime\prime}\right)+(r-2M)\left(\delta\phi_2^{\prime\prime}-\delta C_1^\prime \delta\phi_1^\prime\right)\right)\right]\nonumber\\
& + \frac{1}{4r^2}\left(\frac{G_{4\phi}}{G_4}\right)^2\left[3r(r-2M)\delta\phi_1^{\prime 2}+2\delta\phi_1\left(r(r-2M)\delta\phi_1^{\prime\prime}-2(r-3M)\delta\phi_1^\prime\right)\right]\nonumber\\
& - \frac{1}{2r^2}\frac{G_{4\phi\phi}}{G_4}\left[r(r-2M)\delta\phi_1^{\prime 2}+\delta\phi_1\left(r(r-2M)\delta\phi_1^{\prime\prime}-2(r-3M)\delta\phi_1^\prime\right)\right]\nonumber\\
& - \frac{\alpha_T(r)}{2r^3}\left[-5M+Mr(r-2M)^{-1}-2r(\ell+2)(\ell-1)+r^2(r-2M)\left(\frac{\delta\phi_1^{\prime\prime}}{\delta\phi_1^\prime}\right)^2 + r\left(r(r-2M)\frac{\delta\phi_1^{\prime\prime\prime}}{\delta\phi_1^\prime}-2(r-5M)\frac{\delta\phi_1^{\prime\prime}}{\delta\phi_1^\prime}\right)\right].\label{deltaV2}
\end{align}
\end{subequations}
\end{widetext}
We emphasise that in the above expressions all of the $G_i$ Horndeski functions are evaluated at $\phi=\phi_0$ and $X=0$ (i.e. to zeroth order in the book-keeping parameter $\epsilon$), and as such are \textit{constants}. Note that this approach assumes that the $G_i$ are amenable to an expansion around $\phi=\phi_0$ and $X=0$; this is not the case for Einstein-scalar-Gauss-Bonnet gravity, for example, where the $G_i$ include $\log |X|$ terms \cite{Kobayashi:2011nu}.

As expected, to $\mathcal{O}(\epsilon^0)$ eq.~(\ref{RWhairy}) is simply the well known Regge Wheeler equation describing odd parity gravitational perturbations to a Schwarzschild black hole \cite{Regge:1957td}. 

At $\mathcal{O}(\epsilon^0)$ the effective potential of the Regge-Wheeler equation is modified by $\delta V_1$, which is linear in the first order modifications to the spacetime geometry and scalar profile (and their derivatives). Our expression for $\delta V_1$ with $\delta\phi_1=0$ matches that of eq.~(5.9) in \cite{Franciolini:2018uyq}, which concerns perturbations of hairy black holes in the unitary gauge (i.e. with $\delta\phi_1=0$).

At $\mathcal{O}(\epsilon^2)$, the potential is further modified by $\delta V_2$, which is quadratic in first order `hairy' terms, and linear in the second order modifications. Furthermore, at second order in the perturbative expansion, we see that the frequency term $\omega^2$ is rescaled by a factor of $c_T=1+\epsilon^2\alpha_T$ where $c_T$ is the propagation speed of gravitational waves in Horndeski gravity \cite{DeFelice:2011bh,Bellini:2014fua}. As discussed previously, this term would have to vanish in cosmological settings to satisfy the constraints obtained from GW/GRB170817.

Interestingly, we see that even in the case of pure Schwarzschild geometry (i.e. with $\delta A_i=\delta C_i=0$), if there is non-minimal coupling between the scalar field and metric such that $G_{4\phi}$ or $G_{4\phi\phi} \neq0$, the QNM spectrum can be modified by the presence of a non-trivial scalar radial profile.

With regards to the stability of this slightly hairy black hole, in \cite{Kobayashi:2012kh,Ganguly:2017ort} it is given that a necessary condition to avoid `no-ghost' and `Laplacian' instabilities is that $\mathcal{F},\mathcal{G},$ and $\mathcal{H}$ are all positive. Looking at eq.~(\ref{fgheqs}), we see that for the ansatz given by eq.~(\ref{ansatz}) each of the functions is equal to $2G_4(\phi_0,0)(1+\mathcal{O}(\epsilon))$. As $G_4(\phi_0,0)$ plays the role of the constant background Planck mass in Horndeski, this is clearly positive. Thus each of $\mathcal{F}, \mathcal{G},$ and $\mathcal{H}$ must also be positive for $\epsilon\ll1$.

Turning now to stability against odd parity perturbations, if the potential given by eq.~(\ref{Vgen}) is positive everywhere then we are ensured stability. Using eq.~(\ref{RWhairy}), we see that the potential has the following form:
\begin{align}
\mathscr{V} = A(r)\left(\frac{\ell(\ell+1)}{r^2}-\frac{6M}{r^3}+\mathcal{O}(\epsilon)\right).
\end{align}
Clearly $\mathscr{V}$ vanishes on the horizon $r_H$ where $A(r_H)=0$, and is positive for intermediate values of $r$ (the Regge-Wheeler potential is everywhere positive, and the $\mathcal{O}(\epsilon)$ terms are assumed to be much smaller than the standard GR terms). If each of the functions $\delta A_i$, $\delta C_i$, and $\delta \phi_i$ also decay at least as quickly as $1/r$, then the leading order term of $\mathscr{V}$ is $\ell(\ell+1)/r^2$ as $r\to\infty$ (as in the usual GR case). This ensures that the potential is everywhere positive, including as $r\to\infty$, and can be seen graphically with an example power law ansatz for the hairy functions in figure \ref{fig1}. For other forms of the hairy functions, e.g. those with de Sitter asymptotics, the stability will have to be analysed separately (examples of applying the S-deformation stability analysis to Horndeski theories are given in \cite{Ganguly:2017ort}). 

Eqs.~(\ref{RWhairy})~-~(\ref{deltaVs}) are the main results of this section. In the next section, we will explore how the modifications introduced to eq.~(\ref{RWhairy}) by the small amounts of hair affect the spectrum of QNM frequencies $\omega$ of the black hole. A note of interest, however, is that in the $\omega=0$ limit, eq.~(\ref{RWgen}) could be used to study the tidal deformation of black holes in Horndeski gravity. 

\section{Parameterised QNM Spectrum}\label{QNMsection}

In \cite{Cardoso:2019mqo} (henceforth referred to as Cardoso et al) a formalism is developed such that, given a Schr{\"o}dinger style QNM style equation:
\begin{align}
\left[f(r)\frac{d}{dr}\left(f(r)\frac{d}{dr}\right)+\omega^2-f(r)\tilde{V}\right]\psi=0\label{RWcardoso}
\end{align}
where $f(r)=1-r_H/r$ with $r_H$ the horizon radius, and $\tilde{V}$ is a modified Regge-Wheeler potential in the following form:
\begin{align}
\tilde{V}=\frac{\ell(\ell+1)}{r^2}-\frac{6M}{r^3}+\frac{1}{r_H^2}\sum_{j=0}^{\infty}\alpha_j\left(\frac{r_H}{r}\right)^j,
\end{align}
the spectrum of frequencies $\omega$ can be described in terms of corrections to the standard GR QNM spectrum. The new frequencies are given by:
\begin{align}
\omega = \; \omega_{\,0} + \sum_{j=0}^{\infty}\alpha_j e_j
\end{align}  
where $\omega_{\,0}$ is the unperturbed GR frequency and the $e_j$ are a `basis set' of complex numbers which have been calculated using high precision direct integration of the equations of motion (the reader should consult \cite{Cardoso:2019mqo} for a detailed explanation of this formalism).

We will now use this approach to calculate the modifications to the QNM spectrum induced by the perturbative black hole hair (with an appropriate power law ansatz for the $\delta (A,C,\phi)_i$). For simplicity and compactness we will present results to only first order in the book-keeping parameter $\epsilon$, such that we are seeking the leading order corrections to $\omega$ in the following form:
\begin{align}
\omega=\;\omega_{\,0} + \epsilon \omega_1.
\end{align}
First, however, we must make sure that our eq.~(\ref{RWhairy}) is transformed into the same form as eq.~(\ref{RWcardoso}) so that we can correctly read off the $\alpha_j$ coefficients.

\subsection{Equation manipulation}

To first order in $\epsilon$, the modified Regge-Wheeler equation is given by
\begin{align}
\left[A(r)\frac{d}{dr}\left(A(r)\frac{d}{dr}\right)+\omega^2-A(r)\overline{V}\right]Q=0\label{RWnat}
\end{align}
where the effective potential $\overline{V}$ is given by:
\begin{align}
\overline{V}=\frac{\ell(\ell+1)}{r^2}-\frac{6M}{r^3}+\epsilon\delta V_1\label{Vbar}
\end{align}
Following the procedure introduced in Cardoso et al, the first step to obtain an equation in the form of eq.~(\ref{RWhairy}) is to write:
\begin{align}
A(r)=f(r)Z(r)\label{AtoF}
\end{align}
where $f(r)=1-r_H/r$, and find appropriate expressions for $r_H$ and $Z$ to $O(\epsilon)$. The location of the horizon in our modified spacetime will not be exactly at $r=2M$, but will be corrected due to $\delta A_1$. We thus make the following expansion for the horizon radius:
\begin{align}
r_H=2M+\epsilon \delta r_{H,1}.
\end{align}
To find the new position of the horizon, we require $A(r_H)=0$. Solving order by order in $\epsilon$, we find the following for the location of the horizon:
\begin{align}
\delta r_{H,1}=&\;-2M\delta A_1(2M)\label{deltarH}
\end{align}
with $Z$ thus given by:
\begin{align}
Z(r)=&\; 1+ \epsilon \delta Z_1\nonumber\\
=&\; 1+ \epsilon \frac{\delta A_1(r)-\frac{2M}{r}\delta A_1(2M)}{1-2M/r}
\end{align}
in order to make eq.~(\ref{AtoF}) hold to $O(\epsilon)$.

If we now define $\tilde{Q}=\sqrt{Z}Q$, we transform eq.~(\ref{RWnat}) into
\begin{align}
\left[f(r)\frac{d}{dr}\left(f(r)\frac{d}{dr}\right)+\frac{\omega^2}{Z^2}-f(r)V\right]\tilde{Q}=0\label{RWmod}
\end{align}
where the new potential $V$ is given by:
\begin{align}
V=\frac{\overline{V}}{Z}-\frac{f\left(Z^\prime\right)^2-2Z(fZ^\prime)^\prime}{4Z^2}.
\end{align}
and $\overline{V}$ is still given by eq.~(\ref{Vbar}).

We can expand the $\omega^2$ term in eq.~(\ref{RWmod}) to $O(\epsilon)$ and write it in the following way:
\begin{align}
\frac{\omega^2}{Z^2}=&\;\omega^2(1-2\epsilon\delta Z_1(r_H))-2\epsilon\omega^2(\delta Z_1(r)-\delta Z_1(r_H))\label{omegamod}
\end{align}
The first term on the right hand side of eq.~(\ref{omegamod}) can be seen as a (constant) rescaling of the frequencies. As the second term on the right hand side is already $\mathcal{O}(\epsilon)$, it can be absorbed into the perturbed potential $V$ by setting $\omega=\omega_{\,0}$. This is because any frequency corrections in this term would result in terms $\mathcal{O}(\epsilon^2)$ or higher (which we are neglecting in this section). The final form of the modified Regge Wheeler equation is now in the same form as eq.~(\ref{RWcardoso}):
\begin{align}
\left[f(r)\frac{d}{dr}\left(f(r)\frac{d}{dr}\right)+\tilde{\omega}^2-f(r)\tilde{V}\right]\tilde{Q}=0
\end{align}
where
\begin{align}
\tilde{\omega}^2=&\;\omega^2(1-2\epsilon\delta Z_1(r_H))\\
\tilde{V}=&\;V+\frac{2\epsilon\omega_0^2}{f(r)}(\delta Z_1(r)-\delta Z_1(r_H))\label{Vomega}
\end{align}

The final step before we are able to calculate numerically the modified QNM spectrum of our hairy Horndeski black holes is to assume an appropriate functional form for $\delta A_i$, $\delta C_i$ and $\delta \phi_i$. We will make the following simple power law choices:
\begin{align}
\delta \phi_1 = &\,Q_1\left(\frac{2M}{r}\right),\;\; \delta A_1 = \,a_1\left(\frac{2M}{r}\right)^{2},\;\;\delta C_1 = \,c_{1}\left(\frac{2M}{r}\right)\label{deltaansatz}
\end{align}
so that, in addition to the Horndeski $G_i$ parameters, we have 3 `hairs', $Q_{1}$, $a_{1}$, and $c_{1}$ that can affect our QNM spectrum. Of course the hairy parameters may be related when considering specific solutions, but for now we will assume that they are independent.

With the above ansatz we find the non-zero $\alpha_j$ are given by (absorbing $\epsilon$ into the definitions of $(a,c,Q)_1$):
\begin{subequations}\label{alphas}
\begin{align}
\alpha_0=&\ 8M^2\omega_0^2a_1\\
\alpha_3=&\;\ell(\ell+1)(a_1-c_1)-a_1-2Q_1\frac{G_{4\phi}}{G_4}\\
\alpha_4=&\;\frac{5}{2}\left(a_1+c_1+Q_1\frac{G_{4\phi}}{G_4}\right)
\end{align}
\end{subequations}
leading to the following corrections to the QNM frequency spectrum for the $\ell=2,3$ modes (for example):
\begin{widetext}
\begin{subequations}\label{qnmspec}
\begin{align}
M\omega_1^{\ell=2}=&\;Q_1\frac{G_{4\phi}}{G_4}\left[-0.0126+0.0032i\right]+a_1\left[-0.0267+0.0621i\right]+c_1\left[-0.1296+0.0106i\right]\\
M\omega_1^{\ell=3}=&\;Q_1\frac{G_{4\phi}}{G_4}\left[-0.0075+0.0008i\right]+a_1\left[-0.1326+0.0677i\right]+c_1\left[-0.2040+0.0110i\right]
\end{align}
\end{subequations}
\end{widetext}
where the unperturbed GR frequencies are given by \cite{Berti:2009kk}:
\begin{align}
M\omega_{GR}^{\ell=2}=0.3737-0.0890i,\quad M\omega_{GR}^{\ell=3}=0.5994-0.09270i.
\end{align}

Looking at the real and imaginary parts of the frequencies separately, with $\omega_{\,GR}=\omega_R+i\omega_I$ and $\omega_1=\delta\omega_R + i \delta\omega_I$, we find the following fractional differences for the $\ell=2$ QNM:
\begin{subequations}\label{deltaell2}
\begin{align}
\frac{\delta\omega^{\ell=2}_R}{\omega^{\ell=2}_R} = &\; -(3.37\,\tilde{Q}_1 + 7.14\,a_1 + 34.68\,c_1)\%\\
\frac{\delta\omega^{\ell=2}_I}{\omega^{\ell=2}_I} = &\; -(3.60\,\tilde{Q}_1 + 69.77\,a_1 + 11.91\,c_1)\%,
\end{align}
\end{subequations}
and for the $\ell=3$ mode:
\begin{subequations}\label{deltaell3}
\begin{align}
\frac{\delta\omega^{\ell=3}_R}{\omega^{\ell=3}_R} = &\; -(1.25\,\tilde{Q}_1 + 22.12\,a_1 + 34.03\,c_1)\%\\
\frac{\delta\omega^{\ell=3}_I}{\omega^{\ell=3}_I} = &\; -(0.86\,\tilde{Q}_1 + 73.03\,a_1 + 11.87\,c_1)\%.
\end{align}
\end{subequations}
where $\tilde{Q}_1=Q_1G_{4\phi}/G_4$. 

We see that with, for example\footnote{In \cite{Cardoso:2019mqo} a criterion for the maximum size of any $\alpha_j$ for the perturbative treatment to be valid is given by $|\alpha_j|\ll(1+1/j)^j(j+1)$. An example value of $0.1$ for each of the hairy parameters satisfies this constraint.}, $(a_1, c_1,\tilde{Q}_1)$ all $\sim10^{-1}$, we might expect fractional deviations from the GR spectrum of around $5\%$ and $9\%$ for the real and imaginary parts of the $\ell=2$ frequency, and of $6\%$ and $9\%$ for $\ell=3$. In \cite{2016PhRvL.116v1101A} the least damped QNM of the GW150914 remnant black hole is predicted using the posterior distributions on the mass and spin of the remnant; the predicted fractional uncertainties on the real and imaginary parts of the frequency in that case are around $3\%$ and $8\%$ respectively. Given that in \cite{2016PhRvL.116v1101A} the uncertainties on the real and imaginary parts of the frequency obtained from searching for the QNM in the data are clearly greater than those obtained from using the predicted final mass and spin of the black hole (see Figure 5), the example $\sim5-10\%$ fractional deviations postulated here will likely be dominated by other uncertainties with current detectors.

\begin{figure*}
\caption{Perturbed potential $A(r)\overline{V}(r)$ with $\ell=2$ and $M=1/2$. Each nonzero parameter is given the value $0.05$.}
\label{fig1}
\includegraphics[width=0.9\textwidth]{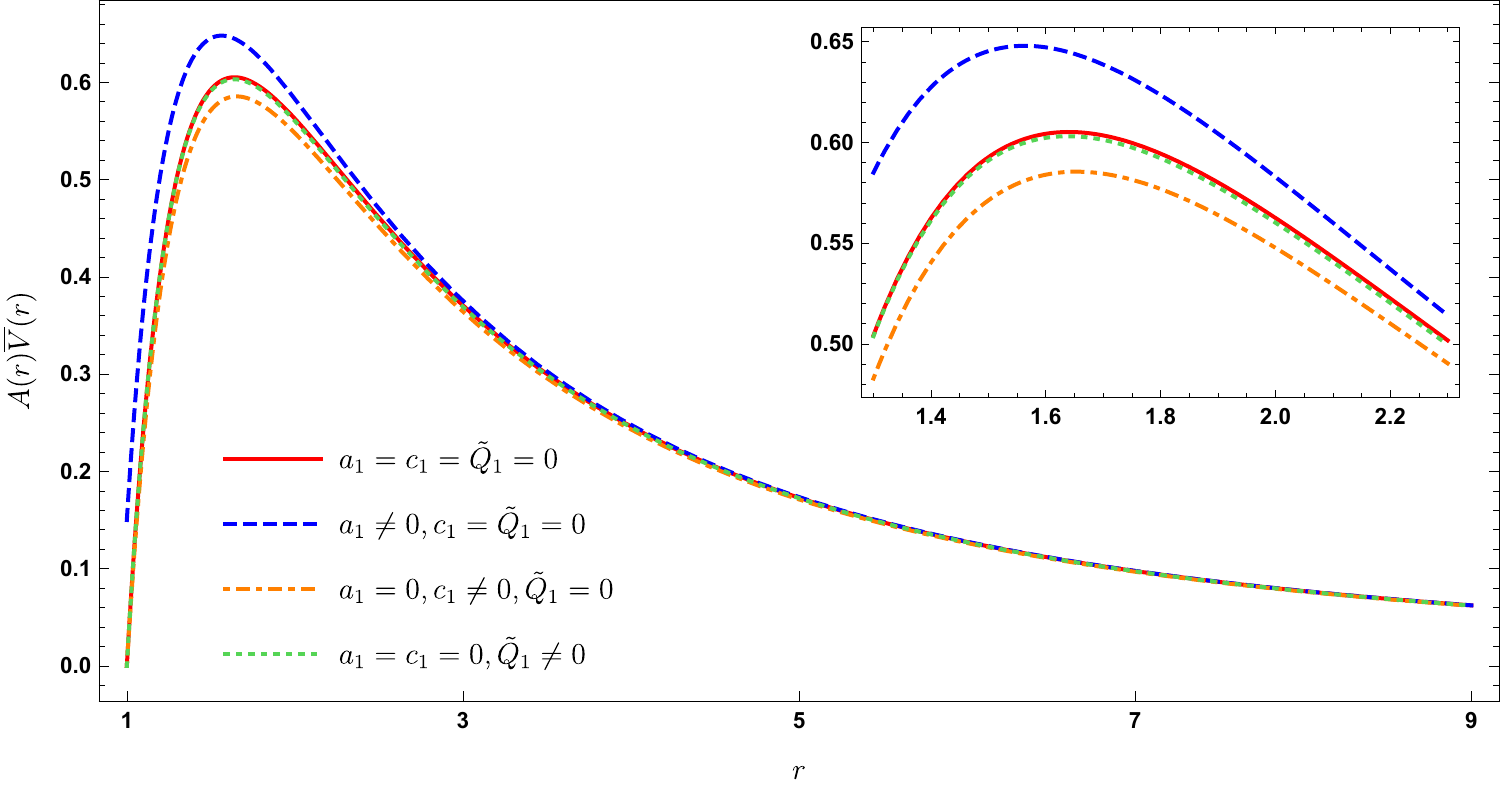}\\
\end{figure*}

Figure \ref{fig1} shows the effect that each of $a_1$, $c_1$, and $\tilde{Q}_1$ has on the form of the first order `hairy' potential $A(r)\overline{V}(r)$ (with $\overline{V}$ given by eq.~(\ref{Vbar})) for $\ell=2$. We see that a non-zero $a_1$ leads to the most noticeable modification to the effective potential. This is unsurprising due to a non-zero $a_1$ leading to a shift in the position of the horizon through eq.~(\ref{deltarH}), as well as giving rise to an effective mass-squared term due to $\alpha_0$ in eq.~(\ref{alphas}). Looking at eq.~(\ref{deltaell2}) - (\ref{deltaell3}), $a_1$ has the most significant effect on $\delta\omega_I$, while $\delta\omega_R$ is most greatly impacted by $c_1$.

The choices made in eq.~(\ref{deltaansatz}) were simply to give a concrete example of a modified QNM in terms of the Horndeski (and new `hairy') parameters; one could of course make a different ansatz of one's choosing to calculate $\omega_1$ (though it should be noted that the numerical results of Cardoso et al only apply for those potentials which can be expressed as a series in inverse integer powers of $r$ - for potentials that do not fit this form alternative methods of calculating the QNM spectrum will have to be deployed \cite{1985RSPSA.402..285L,2009CQGra..26v5003D,Pani:2012zz}). 

Moving beyond the toy model considered above, we could consider more generic forms for $\delta A_1$, $\delta C_1$ and $\delta\phi_1$. For example:
\begin{align}\label{ansatzgeneric}
\delta C_1 =&\,c_1\left(\frac{2M}{r}\right)^m,\;\;\delta A_1 =\,a_1\left(\frac{2M}{r}\right)^n,\nonumber\\
\delta\phi_1=&\,Q_1\left(\frac{2M}{r}\right)^p,
\end{align}
where $m, n,$ and $p$ are positive integers. The contributions to the perturbed potential that these terms lead to are given in appendix \ref{appendixpowerlaw}. In appendix \ref{appendixstealth} we consider the case of a `stealth' black hole, one that has Schwarzschild geometry while being endowed with a non-trivial scalar profile (that is, $\delta A_i = \delta C_i = 0$, $\delta\phi_i \neq0$). We consider both power law and general scalar profiles.

\subsection{Parameter Estimation}\label{paramsec}

We will now follow the Fisher matrix approach of \cite{Berti:2005ys} (to which the reader should refer to for an in-depth treatment of statistical errors and ringdown observations) for performing a parameter estimation analysis on the modified QNM spectrum calculated above. 

In GR, the ringdown signal observed at a gravitational wave detector from a black hole can be modelled as $h=h_+ F_+ + h_\times F_\times$, where $h$ is the total strain, $h_{+,\times}$ is the strain in each of the $+$ and $\times$ polarisations, and $F_{+,\times}$ are pattern functions which depend on the orientation of the detector with respect to the source, and on polarisation angle. In frequency space, the strain in each polarisation is given by:
\begin{align}
h_+=&\;\frac{A^+_{\ell m}}{\sqrt{2}}\left[e^{i\phi_+} S_{\ell m} b_+(f) + e^{-i\phi_+}S^\ast_{\ell m} b_-(f)\right]\\
h_\times=&\;\frac{A^+_{\ell m}N_\times}{\sqrt{2}}\left[e^{i\phi_\times} S_{\ell m} b_+(f) + e^{-i\phi_\times}S^\ast_{\ell m} b_-(f)\right]
\end{align}
where the amplitude $A_+$, amplitude ratio $N_\times$, and phases $\phi_{+,\times}$ are real. The $S$ are complex spin weight 2 spheroidal harmonics, and $b_{\pm}$ are given by:
\begin{align}
b_{\pm}=&\;\frac{1/\tau_{\ell m}}{(1/\tau_{\ell m})^2+4\pi^2(f\pm f_{\ell m})}
\end{align}
where for a given $(\ell, m)$, $\omega_{\ell m}=2\pi f_{\ell m} - i/\tau_{\ell m}$.

In general, Horndeski gravity admits another polarisation of gravitational waves in addition to the usual $+$ and $\times$ polarisations present in GR: that of the `breathing' mode (for a massless scalar field) or mixed breathing-longitudinal mode (for a massive field) \cite{Hou:2017bqj,Gong:2018ybk}. This mode is associated with the oscillations of the additional scalar degree of freedom. As we are only interested in considering the effect of black hole hair on the frequency spectrum associated with the odd parity gravitational degree of freedom in this paper, and not of the scalar degree of freedom, we will neglect the additional polarisation. This is equivalent to setting the scalar field perturbation to zero, and focussing solely on the gravitational perturbations. Thus we will model the strain $h$ as the usual sum of the GR polarisations in order to gain an estimate of how well one could constrain deviations from the GR QNM spectrum. In \cite{Martel:2005ir} it is shown that the odd parity metric degree of freedom contributes to both $+$ and $\times$ polarisations. 

We are interested in calculating the statistical errors in determining the `hairy' parameters that affect the QNM spectrum, and as such we assume that the mass $M$ of the black hole, and thus the unperturbed QNM frequency $\omega_{\,0}$, is known. Furthermore, we will assume that $N_\times=1$ and $\phi_+=\phi_\times=0$, and that $A_+$ is known (effectively resulting in us fixing a specific signal-noise-ratio $\rho$). In \cite{Berti:2005ys} it is shown that the results for statistical errors are not strongly affected by the values of $N_\times$ or of the phases.

Furthermore we will assume that $c_1=0$, thus we are effectively considering a Reissner-Nordstrom-like black hole with a $1/r$ scalar profile. Again with $\tilde{Q}_1=Q_1G_{4\phi}/G_4$, and absorbing the book-keeping parameter $\epsilon$ into the definition of our hairy parameters (such that $\epsilon a_1\rightarrow a_1$ etc), we can write the oscillation frequency and damping time of the perturbed $\ell=2$ mode (for example) as follows:
\begin{subequations}
\begin{align}
2\pi f=&\;2\pi f_0-0.0126\tilde{Q}_1/M-0.0267a_1/M\label{f2}\\
\tau^{-1}=&\;\tau_0^{-1}-0.0032\tilde{Q}_1/M-0.0621a_1/M.\label{tau2}
\end{align}
\end{subequations}

Using the Fisher matrix formalism laid out in \cite{Berti:2005ys}, and remembering that we are assuming $M$ to be known exactly, we calculate the following errors for $\tilde{Q}_1$ and $a_1$:
\begin{subequations}
\begin{align}
\sigma^2_{\tilde{Q}_1}=&\;\frac{1}{2\rho^2q^2}\frac{f^{\prime 2}q^2(1+4q^2)-2fqf^\prime q^\prime + f^2q^{\prime 2}}{\left(\dot{f}q^\prime-f^\prime \dot{q}\right)^2}\\
\sigma^2_{a_1}=&\;\frac{1}{2\rho^2q^2}\frac{\dot{f}^2q^2(1+4q^2)-2fq\dot{f}\dot{q} + f^2\dot{q}^2}{\left(\dot{f}q^\prime-f^\prime \dot{q}\right)^2}
\end{align}
\end{subequations}
where $q=\pi f \tau$ is the `quality factor' of a given oscillation mode, and we now use the notation $F^\prime\equiv\frac{\partial F}{\partial a_1}$ and $\dot{F}\equiv\frac{\partial F}{\partial \tilde{Q}_1}$ for a quantity $F$.

Assuming a detection of the $\ell=2$ mode, we can use the expressions given in eq.~(\ref{f2}) and (\ref{tau2}) to calculate the errors. Additionally setting $\tilde{Q}_1=a_1=0$, we interpret the following errors as `detectability' limits on the parameters:
\begin{align}
\rho\sigma_{\tilde{Q}_1}\approx&\;12,\quad\rho\sigma_{a_1}\approx\;2.
\end{align}
With an SNR of $\rho\sim10^2$, which could be typical of LISA events, we thus have that $\sigma_{\tilde{Q}_1}\approx0.1$ whilst $\sigma_{a_1}\approx0.02$ (assuming that the mass of the black hole is known with absolute precision). As discussed previously with regards to GW150914, however, this is clearly a highly optimistic scenario for current detectors.

Eq.~(\ref{deltaell2}) shows that the metric Reissner-Nordstrom-like hair $a_1$ has a more significant effect on the frequency and damping time than the scalar hair $\tilde{Q}_1$, so it is unsurprising that we find it possible to constrain $a_1$ to a greater degree than the scalar hair. In fact, in general it perhaps makes intuitive sense that the odd parity QNMs are more affected by modifications to the spacetime geometry than to the scalar profile, given that the scalar perturbations only couple to the even parity sector of the gravitational perturbations.

\section{Discussion}\label{discussion}

In this paper we have studied the QNMs associated with odd parity gravitational perturbations of spherically symmetric black holes in Horndeski gravity. By assuming that the background solutions for the spacetime geometry and Horndeski scalar field are well described to first order by the hairless Schwarzschild solution, we can treat the effect of any black hole `hair' as small modifications. 

Making use of the results for generic spherically symmetric black holes derived in \cite{Ganguly:2017ort}, we present a modified Regge-Wheeler style equation, eq.~(\ref{RWhairy}), describing odd parity gravitational perturbations. Eq.~(\ref{RWhairy}) takes into account effects induced by \textit{generic} modifications to both the spacetime geometry and to the background radial profile of the Horndeski scalar field. Labelling the background modifications by a book-keeping parameter $\epsilon$ to keep track of the order of `smallness', we present results to $\mathcal{O}(\epsilon^2)$. We show that the odd parity perturbations are not only affected by changes to the background spacetime, but also by the scalar field profile, with the `nonminimal' and `derivative' couplings to curvature $G_4$ and $G_5$ in the Horndeski action playing a role in eq.~(\ref{RWhairy}).

Through the formalism of \cite{Cardoso:2019mqo} the odd parity QNM spectrum of such perturbatively hairy black holes can be calculated (assuming an inverse power law ansatz for the modifications to both the spacetime and scalar profile). In eq.~(\ref{qnmspec}) the first order modifications to the QNM spectrum are presented for the $\ell=2,3$ modes for a specific power law model of the black hole hair. It is straightforward to calculate the modifications for other $\ell$ using the results of this paper combined with the numerical data provided in \cite{Cardoso:2019mqo}. Results for more generic models of the black hole hair and scalar profile are provided in appendices \ref{appendixpowerlaw} and \ref{appendixstealth}.

We have thus presented a straightforward way to associate deviations from the expected GR QNM spectrum of black holes to not only modifications to the background spacetime, but also to fundamental parameters of a modified gravity theory (in this case, the $G_i$ of Horndeski gravity). In section \ref{paramsec} we perform a simple parameter estimation exercise based on the hypothetical observation of the $\ell=2$ QNM of a black hole whose mass we are assuming to know. With SNRs typical of LISA detections we show that the `hairy' black hole parameters introduced could potentially be well constrained. The predicted fractional deviations from the GR frequency spectrum are likely to be dominated by other sources of uncertainty when considering current detectors, however.

There are of course numerous ways to develop the work presented here. As mentioned briefly in section \ref{perturbsec}, eq.~(\ref{RWgen}) with $\omega=0$ could be used to study the tidal deformations of black holes in Horndeski gravity. Furthermore, one could attempt to find exact solutions for the $\delta(A,C,\phi)_i$ in different realisations of Horndeski gravity (through finding `order-by-order' solutions of otherwise). 

The most natural extension to this work is of course to study the even parity sector of perturbations in Horndeski gravity. In general the even parity sector of gravitational perturbations is more complex than the odd parity sector, and in Horndeski gravity this is only further complicated through the coupling of scalar perturbations to the gravitational modes. The formalism of \cite{Cardoso:2019mqo} has, usefully, been expanded to apply to coupled QNM equations in \cite{McManus:2019ulj}, thus calculating the modified even parity QNM spectrum should be relatively straightforward once the relevant equations have been derived. Such an analysis will then provide a complete description of `almost' Schwarzschild QNMs in Horndeski theory. 

Perhaps the most important extension to this line of research is to include black hole spin, given that the black holes currently observed through merger events appear to possess non-negligible angular momentum \cite{LIGOScientific:2018mvr}. As a first step, one could consider studying slowly rotating `almost Kerr' black holes in Horndeski gravity by introducing another `hairy' function in the $g_{t\phi}$ component of the slowly rotating Kerr metric. More ambitiously, perhaps an `almost' Teukoslky like equation could be found by introducing perturbations to the full Kerr solution. Perturbations of a stealth Kerr black hole (i.e. a Kerr geometry endowed with a non-trivial scalar profile) in Degenerate Higher Order Scalar Tensor theories have been studied in \cite{Charmousis:2019fre,Charmousis:2019vnf}, and of a Kerr black hole in $f(R)$ gravity in \cite{Suvorov:2019qow}.

\section*{Acknowledgments}
\vspace{-0.2in}
\noindent OJT would like to thank Pedro Ferreira, Vitor Cardoso, and Adrien Kuntz for useful conversations whilst preparing this work. OJT acknowledges support from the European Research Council Grant No: 693024.

\appendix

\section{Second-order action coefficients}\label{appendixaction}

The coefficients $a_i$ of the second order Horndeski action given by eq.~(\ref{secondorderaction}) were found in \cite{Kobayashi:2012kh,Ganguly:2017ort} to be:
\begin{subequations}
\begin{align}
a_1=&\,\frac{\ell(\ell+1)}{4C}\left[\left(C^\prime\sqrt{\frac{B}{A}}\mathcal{H}\right)^\prime+\frac{(\ell-1)(\ell+2)}{\sqrt{AB}}\mathcal{F}\right]\\
a_2=&\,-\frac{\ell(\ell+1)(\ell-1)(\ell+2)}{4C}\sqrt{AB}\,\mathcal{G}\\
a_3=&\;\frac{\ell(\ell+1)}{4}\sqrt{\frac{B}{A}}\mathcal{H}
\end{align}
\end{subequations}
where $\mathcal{F},\mathcal{G},\mathcal{H}$ are given by eq.~(\ref{fgheqs}).

\section{General power law model}\label{appendixpowerlaw}

For the hairy black hole model given by eq.~(\ref{ansatzgeneric}), we find the following first order contributions to the effective potential $\tilde{V}$:
\begin{widetext}
\begin{align}
(2M)^2\tilde{V}_1 =& \, \tilde{Q}_1\left(\frac{2M}{r}\right)^{p+2}\frac{p}{2}\left[\left(\frac{2M}{r}\right)(p+4)-(p+3)\right]+ c_1\left(\frac{2M}{r}\right)^{m+2}\frac{1}{2}\left[\left(\frac{2M}{r}\right)m(m+4)-(m+4)(m-1)-2\ell(\ell+1)\right]\nonumber\\
&+\frac{a_1}{(1-2 M/r)^2}\frac{1}{2}\left(\frac{2M}{r}\right)^2 \left[\left(\frac{2M}{r}\right) \left(2\left(\ell^2+\ell-1\right)-(2 \ell (\ell+1)+5) \left(\frac{2M}{r}\right)+6 \left(\frac{2M}{r}\right)^2\right)\right.\nonumber\\
&\left.+
   \left(\frac{2M}{r}\right)^n \left(\left(\frac{2M}{r}\right) (2 \ell (\ell+1)-n (2 n+5))+ (-2 \ell (\ell+1)+n (n+3)+4)+\left(\frac{2M}{r}\right)^2 (n-1) (n+3)+4 r^2
   \omega _0^2\right)\right.\nonumber\\
   &\left.+4 r^2 \omega _0^2\left(n\left(1-\frac{2M}{r}\right)-1\right)\right]
\end{align}
\end{widetext}
where $\tilde{Q}_1=Q_1G_{4\phi}/G_4$, and $\tilde{V}_1$ denotes the linear in $\epsilon$ component of $\tilde{V}$. 

We see that for arbitrary $n$ it is non-trivial to find the $\alpha_j$ that contribute to the modifications to the QNM spectrum because of the $(1-2M/r)^2$ factor in the denominator of the $a_1$ term. For $0<n<6$, however, the factor in the denominator cancels with the numerator, leading to a finite number of terms in powers of $1/r$. In cases where the factor does not cancel, one would have to Taylor expand $(1-2M/r)^{-2}$ to an appropriate number of terms until the desired `accuracy' of the modification to the QNM frequency is reached (i.e. until adding additional terms from the Taylor expansion changes $\omega_1$ less than some predetermined threshold).

\section{Stealth black holes}\label{appendixstealth}

We will now study the special case of a `stealth' black hole - i.e. a black hole endowed with non trivial scalar hair whilst maintaining Schwarzschild geometry (and therefore with $\delta A_i = \delta C_i = 0$). Working to linear order in the hair terms, which in this case is just $\delta\phi_1$, we find the following corrections to the $\ell=2$ QNM using the $1/(\ell+1/2)$ expansion method of \cite{2009CQGra..26v5003D} when the scalar profile is of the form given in eq.~(\ref{ansatzgeneric}):
\begin{widetext}
\begin{subequations}\label{phigenericqnm}
\begin{align}
\frac{M\delta\omega_R^{\ell=2}}{Q_1G_{4\phi}/G_4} = & -\frac{p(2/3)^p}{2125764000000\sqrt{3}}\left(1215 p^9-68445 p^8+1004850 p^7-338634 p^6-73083789 p^5+416702775 p^4\right.\nonumber\\
   &\left.+398808500
   p^3-7606822416 p^2+27804991944 p+49508814560\right)\\
\frac{M\delta\omega_I^{\ell=2}}{Q_1G_{4\phi}/G_4} = & \frac{p(2/3)^p}{191318760000000\sqrt{3}}\left(-729 p^{11}+60426 p^{10}-1503495 p^9+8335332 p^8+124548741
   p^7-1481521014 p^6\right.\nonumber\\
   &\left.+2350263795 p^5+36358262708 p^4-223133829602
   p^3+273476762848 p^2+2199852417910 p\right.\nonumber\\
   &\left.-811320378640\right).
\end{align}
\end{subequations}
\end{widetext}
For $m=1$ the above expressions give $M\delta\omega_R^{\ell=2}=-0.0128(Q_1G_{4\phi}/G_4)$ and $M\delta\omega_I^{\ell=2}=0.0030(Q_1G_{4\phi}/G_4)$, in good agreement eq.~(\ref{qnmspec}) which utilised the numerical results of Cardoso et al. 

As $\delta V_1$ is linear in the scalar profile $\delta\phi_1$, and because we are considering only linear corrections to the QNM frequencies, one can use eq.~(\ref{phigenericqnm}) to construct the corrected $\ell=2$ mode for a scalar profile of the following form:
\begin{align}
\delta\phi_1(r) = \sum_{p_i=1}^\infty Q_1^{(p_i)}\left(\frac{2M}{r}\right)^{p_i}
\end{align}
where each of the $Q_1^{(p_i)}\ll1$ but of similar magnitude to each other. One would simply use eq.~(\ref{phigenericqnm}) to calculate the $\delta\omega_{R,I}$ for each value of $p_i$ required and then sum the corrections to find the final frequency.

For a generic scalar profile (i.e. not limited to integer power laws), we have calculated the QNM frequencies as a function of $\ell$ and $\delta\phi_1$ using the expansion method of \cite{2009CQGra..26v5003D}. These include very lengthy expressions involving up to 12 derivatives of $\delta\phi_1$ so we do not include them here, but they are presented in a Mathematica notebook available online \cite{oxwebsite}. Indeed, the expressions given in eq.~(\ref{phigenericqnm}) are in fact valid for any $p$.

\bibliography{RefModifiedGravity}

\begin{thebibliography}{91}%
\makeatletter
\providecommand \@ifxundefined [1]{%
 \@ifx{#1\undefined}
}%
\providecommand \@ifnum [1]{%
 \ifnum #1\expandafter \@firstoftwo
 \else \expandafter \@secondoftwo
 \fi
}%
\providecommand \@ifx [1]{%
 \ifx #1\expandafter \@firstoftwo
 \else \expandafter \@secondoftwo
 \fi
}%
\providecommand \natexlab [1]{#1}%
\providecommand \enquote  [1]{``#1''}%
\providecommand \bibnamefont  [1]{#1}%
\providecommand \bibfnamefont [1]{#1}%
\providecommand \citenamefont [1]{#1}%
\providecommand \href@noop [0]{\@secondoftwo}%
\providecommand \href [0]{\begingroup \@sanitize@url \@href}%
\providecommand \@href[1]{\@@startlink{#1}\@@href}%
\providecommand \@@href[1]{\endgroup#1\@@endlink}%
\providecommand \@sanitize@url [0]{\catcode `\\12\catcode `\$12\catcode
  `\&12\catcode `\#12\catcode `\^12\catcode `\_12\catcode `\%12\relax}%
\providecommand \@@startlink[1]{}%
\providecommand \@@endlink[0]{}%
\providecommand \url  [0]{\begingroup\@sanitize@url \@url }%
\providecommand \@url [1]{\endgroup\@href {#1}{\urlprefix }}%
\providecommand \urlprefix  [0]{URL }%
\providecommand \Eprint [0]{\href }%
\providecommand \doibase [0]{http://dx.doi.org/}%
\providecommand \selectlanguage [0]{\@gobble}%
\providecommand \bibinfo  [0]{\@secondoftwo}%
\providecommand \bibfield  [0]{\@secondoftwo}%
\providecommand \translation [1]{[#1]}%
\providecommand \BibitemOpen [0]{}%
\providecommand \bibitemStop [0]{}%
\providecommand \bibitemNoStop [0]{.\EOS\space}%
\providecommand \EOS [0]{\spacefactor3000\relax}%
\providecommand \BibitemShut  [1]{\csname bibitem#1\endcsname}%
\let\auto@bib@innerbib\@empty
\bibitem [{\citenamefont {Abbott}\ \emph {et~al.}(2018)\citenamefont {Abbott}
  \emph {et~al.}}]{LIGOScientific:2018mvr}%
  \BibitemOpen
  \bibfield  {author} {\bibinfo {author} {\bibfnamefont {B.~P.}\ \bibnamefont
  {Abbott}} \emph {et~al.} (\bibinfo {collaboration} {LIGO Scientific,
  Virgo}),\ }\href@noop {} {\  (\bibinfo {year} {2018})},\ \Eprint
  {http://arxiv.org/abs/1811.12907} {arXiv:1811.12907 [astro-ph.HE]}
  \BibitemShut {NoStop}%
\bibitem [{\citenamefont {Dreyer}\ \emph {et~al.}(2004)\citenamefont {Dreyer},
  \citenamefont {Kelly}, \citenamefont {Krishnan}, \citenamefont {Finn},
  \citenamefont {Garrison},\ and\ \citenamefont
  {Lopez-Aleman}}]{Dreyer:2003bv}%
  \BibitemOpen
  \bibfield  {author} {\bibinfo {author} {\bibfnamefont {O.}~\bibnamefont
  {Dreyer}}, \bibinfo {author} {\bibfnamefont {B.~J.}\ \bibnamefont {Kelly}},
  \bibinfo {author} {\bibfnamefont {B.}~\bibnamefont {Krishnan}}, \bibinfo
  {author} {\bibfnamefont {L.~S.}\ \bibnamefont {Finn}}, \bibinfo {author}
  {\bibfnamefont {D.}~\bibnamefont {Garrison}}, \ and\ \bibinfo {author}
  {\bibfnamefont {R.}~\bibnamefont {Lopez-Aleman}},\ }\href {\doibase
  10.1088/0264-9381/21/4/003} {\bibfield  {journal} {\bibinfo  {journal}
  {Class. Quant. Grav.}\ }\textbf {\bibinfo {volume} {21}},\ \bibinfo {pages}
  {787} (\bibinfo {year} {2004})},\ \Eprint
  {http://arxiv.org/abs/gr-qc/0309007} {arXiv:gr-qc/0309007 [gr-qc]}
  \BibitemShut {NoStop}%
\bibitem [{\citenamefont {Berti}\ \emph {et~al.}(2006)\citenamefont {Berti},
  \citenamefont {Cardoso},\ and\ \citenamefont {Will}}]{Berti:2005ys}%
  \BibitemOpen
  \bibfield  {author} {\bibinfo {author} {\bibfnamefont {E.}~\bibnamefont
  {Berti}}, \bibinfo {author} {\bibfnamefont {V.}~\bibnamefont {Cardoso}}, \
  and\ \bibinfo {author} {\bibfnamefont {C.~M.}\ \bibnamefont {Will}},\ }\href
  {\doibase 10.1103/PhysRevD.73.064030} {\bibfield  {journal} {\bibinfo
  {journal} {Phys. Rev.}\ }\textbf {\bibinfo {volume} {D73}},\ \bibinfo {pages}
  {064030} (\bibinfo {year} {2006})},\ \Eprint
  {http://arxiv.org/abs/gr-qc/0512160} {arXiv:gr-qc/0512160 [gr-qc]}
  \BibitemShut {NoStop}%
\bibitem [{\citenamefont {Gossan}\ \emph {et~al.}(2012)\citenamefont {Gossan},
  \citenamefont {Veitch},\ and\ \citenamefont {Sathyaprakash}}]{Gossan:2011ha}%
  \BibitemOpen
  \bibfield  {author} {\bibinfo {author} {\bibfnamefont {S.}~\bibnamefont
  {Gossan}}, \bibinfo {author} {\bibfnamefont {J.}~\bibnamefont {Veitch}}, \
  and\ \bibinfo {author} {\bibfnamefont {B.~S.}\ \bibnamefont
  {Sathyaprakash}},\ }\href {\doibase 10.1103/PhysRevD.85.124056} {\bibfield
  {journal} {\bibinfo  {journal} {Phys. Rev.}\ }\textbf {\bibinfo {volume}
  {D85}},\ \bibinfo {pages} {124056} (\bibinfo {year} {2012})},\ \Eprint
  {http://arxiv.org/abs/1111.5819} {arXiv:1111.5819 [gr-qc]} \BibitemShut
  {NoStop}%
\bibitem [{\citenamefont {Meidam}\ \emph {et~al.}(2014)\citenamefont {Meidam},
  \citenamefont {Agathos}, \citenamefont {Van Den~Broeck}, \citenamefont
  {Veitch},\ and\ \citenamefont {Sathyaprakash}}]{Meidam:2014jpa}%
  \BibitemOpen
  \bibfield  {author} {\bibinfo {author} {\bibfnamefont {J.}~\bibnamefont
  {Meidam}}, \bibinfo {author} {\bibfnamefont {M.}~\bibnamefont {Agathos}},
  \bibinfo {author} {\bibfnamefont {C.}~\bibnamefont {Van Den~Broeck}},
  \bibinfo {author} {\bibfnamefont {J.}~\bibnamefont {Veitch}}, \ and\ \bibinfo
  {author} {\bibfnamefont {B.~S.}\ \bibnamefont {Sathyaprakash}},\ }\href
  {\doibase 10.1103/PhysRevD.90.064009} {\bibfield  {journal} {\bibinfo
  {journal} {Phys. Rev.}\ }\textbf {\bibinfo {volume} {D90}},\ \bibinfo {pages}
  {064009} (\bibinfo {year} {2014})},\ \Eprint {http://arxiv.org/abs/1406.3201}
  {arXiv:1406.3201 [gr-qc]} \BibitemShut {NoStop}%
\bibitem [{\citenamefont {Berti}\ \emph {et~al.}(2016)\citenamefont {Berti},
  \citenamefont {Sesana}, \citenamefont {Barausse}, \citenamefont {Cardoso},\
  and\ \citenamefont {Belczynski}}]{Berti:2016lat}%
  \BibitemOpen
  \bibfield  {author} {\bibinfo {author} {\bibfnamefont {E.}~\bibnamefont
  {Berti}}, \bibinfo {author} {\bibfnamefont {A.}~\bibnamefont {Sesana}},
  \bibinfo {author} {\bibfnamefont {E.}~\bibnamefont {Barausse}}, \bibinfo
  {author} {\bibfnamefont {V.}~\bibnamefont {Cardoso}}, \ and\ \bibinfo
  {author} {\bibfnamefont {K.}~\bibnamefont {Belczynski}},\ }\href {\doibase
  10.1103/PhysRevLett.117.101102} {\bibfield  {journal} {\bibinfo  {journal}
  {Phys. Rev. Lett.}\ }\textbf {\bibinfo {volume} {117}},\ \bibinfo {pages}
  {101102} (\bibinfo {year} {2016})},\ \Eprint
  {http://arxiv.org/abs/1605.09286} {arXiv:1605.09286 [gr-qc]} \BibitemShut
  {NoStop}%
\bibitem [{\citenamefont {Berti}\ \emph {et~al.}(2018)\citenamefont {Berti},
  \citenamefont {Yagi}, \citenamefont {Yang},\ and\ \citenamefont
  {Yunes}}]{Berti:2018vdi}%
  \BibitemOpen
  \bibfield  {author} {\bibinfo {author} {\bibfnamefont {E.}~\bibnamefont
  {Berti}}, \bibinfo {author} {\bibfnamefont {K.}~\bibnamefont {Yagi}},
  \bibinfo {author} {\bibfnamefont {H.}~\bibnamefont {Yang}}, \ and\ \bibinfo
  {author} {\bibfnamefont {N.}~\bibnamefont {Yunes}},\ }\href@noop {} {\
  (\bibinfo {year} {2018})},\ \Eprint {http://arxiv.org/abs/1801.03587}
  {arXiv:1801.03587 [gr-qc]} \BibitemShut {NoStop}%
\bibitem [{\citenamefont {Baibhav}\ and\ \citenamefont
  {Berti}(2019)}]{Baibhav:2018rfk}%
  \BibitemOpen
  \bibfield  {author} {\bibinfo {author} {\bibfnamefont {V.}~\bibnamefont
  {Baibhav}}\ and\ \bibinfo {author} {\bibfnamefont {E.}~\bibnamefont
  {Berti}},\ }\href {\doibase 10.1103/PhysRevD.99.024005} {\bibfield  {journal}
  {\bibinfo  {journal} {Phys. Rev.}\ }\textbf {\bibinfo {volume} {D99}},\
  \bibinfo {pages} {024005} (\bibinfo {year} {2019})},\ \Eprint
  {http://arxiv.org/abs/1809.03500} {arXiv:1809.03500 [gr-qc]} \BibitemShut
  {NoStop}%
\bibitem [{\citenamefont {Giesler}\ \emph {et~al.}(2019)\citenamefont
  {Giesler}, \citenamefont {Isi}, \citenamefont {Scheel},\ and\ \citenamefont
  {Teukolsky}}]{Giesler:2019uxc}%
  \BibitemOpen
  \bibfield  {author} {\bibinfo {author} {\bibfnamefont {M.}~\bibnamefont
  {Giesler}}, \bibinfo {author} {\bibfnamefont {M.}~\bibnamefont {Isi}},
  \bibinfo {author} {\bibfnamefont {M.}~\bibnamefont {Scheel}}, \ and\ \bibinfo
  {author} {\bibfnamefont {S.}~\bibnamefont {Teukolsky}},\ }\href@noop {} {\
  (\bibinfo {year} {2019})},\ \Eprint {http://arxiv.org/abs/1903.08284}
  {arXiv:1903.08284 [gr-qc]} \BibitemShut {NoStop}%
\bibitem [{\citenamefont {Bhagwat}\ \emph
  {et~al.}(2019{\natexlab{a}})\citenamefont {Bhagwat}, \citenamefont {Forteza},
  \citenamefont {Pani},\ and\ \citenamefont {Ferrari}}]{Bhagwat:2019dtm}%
  \BibitemOpen
  \bibfield  {author} {\bibinfo {author} {\bibfnamefont {S.}~\bibnamefont
  {Bhagwat}}, \bibinfo {author} {\bibfnamefont {X.~J.}\ \bibnamefont
  {Forteza}}, \bibinfo {author} {\bibfnamefont {P.}~\bibnamefont {Pani}}, \
  and\ \bibinfo {author} {\bibfnamefont {V.}~\bibnamefont {Ferrari}},\
  }\href@noop {} {\  (\bibinfo {year} {2019}{\natexlab{a}})},\ \Eprint
  {http://arxiv.org/abs/1910.08708} {arXiv:1910.08708 [gr-qc]} \BibitemShut
  {NoStop}%
\bibitem [{\citenamefont {Bhagwat}\ \emph
  {et~al.}(2019{\natexlab{b}})\citenamefont {Bhagwat}, \citenamefont {Cabero},
  \citenamefont {Capano}, \citenamefont {Krishnan},\ and\ \citenamefont
  {Brown}}]{Bhagwat:2019bwv}%
  \BibitemOpen
  \bibfield  {author} {\bibinfo {author} {\bibfnamefont {S.}~\bibnamefont
  {Bhagwat}}, \bibinfo {author} {\bibfnamefont {M.}~\bibnamefont {Cabero}},
  \bibinfo {author} {\bibfnamefont {C.~D.}\ \bibnamefont {Capano}}, \bibinfo
  {author} {\bibfnamefont {B.}~\bibnamefont {Krishnan}}, \ and\ \bibinfo
  {author} {\bibfnamefont {D.~A.}\ \bibnamefont {Brown}},\ }\href@noop {} {\
  (\bibinfo {year} {2019}{\natexlab{b}})},\ \Eprint
  {http://arxiv.org/abs/1910.13203} {arXiv:1910.13203 [gr-qc]} \BibitemShut
  {NoStop}%
\bibitem [{\citenamefont {Maselli}\ \emph {et~al.}(2019)\citenamefont
  {Maselli}, \citenamefont {Pani}, \citenamefont {Gualtieri},\ and\
  \citenamefont {Berti}}]{Maselli:2019mjd}%
  \BibitemOpen
  \bibfield  {author} {\bibinfo {author} {\bibfnamefont {A.}~\bibnamefont
  {Maselli}}, \bibinfo {author} {\bibfnamefont {P.}~\bibnamefont {Pani}},
  \bibinfo {author} {\bibfnamefont {L.}~\bibnamefont {Gualtieri}}, \ and\
  \bibinfo {author} {\bibfnamefont {E.}~\bibnamefont {Berti}},\ }\href@noop {}
  {\  (\bibinfo {year} {2019})},\ \Eprint {http://arxiv.org/abs/1910.12893}
  {arXiv:1910.12893 [gr-qc]} \BibitemShut {NoStop}%
\bibitem [{\citenamefont {Ota}\ and\ \citenamefont
  {Chirenti}(2019)}]{Ota:2019bzl}%
  \BibitemOpen
  \bibfield  {author} {\bibinfo {author} {\bibfnamefont {I.}~\bibnamefont
  {Ota}}\ and\ \bibinfo {author} {\bibfnamefont {C.}~\bibnamefont {Chirenti}},\
  }\href@noop {} {\  (\bibinfo {year} {2019})},\ \Eprint
  {http://arxiv.org/abs/1911.00440} {arXiv:1911.00440 [gr-qc]} \BibitemShut
  {NoStop}%
\bibitem [{\citenamefont {Cabero}\ \emph {et~al.}(2019)\citenamefont {Cabero},
  \citenamefont {Westerweck}, \citenamefont {Capano}, \citenamefont {Kumar},
  \citenamefont {Nielsen},\ and\ \citenamefont {Krishnan}}]{Cabero:2019zyt}%
  \BibitemOpen
  \bibfield  {author} {\bibinfo {author} {\bibfnamefont {M.}~\bibnamefont
  {Cabero}}, \bibinfo {author} {\bibfnamefont {J.}~\bibnamefont {Westerweck}},
  \bibinfo {author} {\bibfnamefont {C.~D.}\ \bibnamefont {Capano}}, \bibinfo
  {author} {\bibfnamefont {S.}~\bibnamefont {Kumar}}, \bibinfo {author}
  {\bibfnamefont {A.~B.}\ \bibnamefont {Nielsen}}, \ and\ \bibinfo {author}
  {\bibfnamefont {B.}~\bibnamefont {Krishnan}},\ }\href@noop {} {\  (\bibinfo
  {year} {2019})},\ \Eprint {http://arxiv.org/abs/1911.01361} {arXiv:1911.01361
  [gr-qc]} \BibitemShut {NoStop}%
\bibitem [{\citenamefont {{Chandrasekhar}}(1975)}]{1975RSPSA.343..289C}%
  \BibitemOpen
  \bibfield  {author} {\bibinfo {author} {\bibfnamefont {S.}~\bibnamefont
  {{Chandrasekhar}}},\ }\href {\doibase 10.1098/rspa.1975.0066} {\bibfield
  {journal} {\bibinfo  {journal} {Proceedings of the Royal Society of London
  Series A}\ }\textbf {\bibinfo {volume} {343}},\ \bibinfo {pages} {289}
  (\bibinfo {year} {1975})}\BibitemShut {NoStop}%
\bibitem [{\citenamefont {Nollert}(1999)}]{0264-9381-16-12-201}%
  \BibitemOpen
  \bibfield  {author} {\bibinfo {author} {\bibfnamefont {H.-P.}\ \bibnamefont
  {Nollert}},\ }\href {http://stacks.iop.org/0264-9381/16/i=12/a=201}
  {\bibfield  {journal} {\bibinfo  {journal} {Classical and Quantum Gravity}\
  }\textbf {\bibinfo {volume} {16}},\ \bibinfo {pages} {R159} (\bibinfo {year}
  {1999})}\BibitemShut {NoStop}%
\bibitem [{\citenamefont {Kokkotas}\ and\ \citenamefont
  {Schmidt}(1999)}]{Kokkotas:1999bd}%
  \BibitemOpen
  \bibfield  {author} {\bibinfo {author} {\bibfnamefont {K.~D.}\ \bibnamefont
  {Kokkotas}}\ and\ \bibinfo {author} {\bibfnamefont {B.~G.}\ \bibnamefont
  {Schmidt}},\ }\href {\doibase 10.12942/lrr-1999-2} {\bibfield  {journal}
  {\bibinfo  {journal} {Living Rev. Rel.}\ }\textbf {\bibinfo {volume} {2}},\
  \bibinfo {pages} {2} (\bibinfo {year} {1999})},\ \Eprint
  {http://arxiv.org/abs/gr-qc/9909058} {arXiv:gr-qc/9909058 [gr-qc]}
  \BibitemShut {NoStop}%
\bibitem [{\citenamefont {Berti}\ \emph {et~al.}(2009)\citenamefont {Berti},
  \citenamefont {Cardoso},\ and\ \citenamefont {Starinets}}]{Berti:2009kk}%
  \BibitemOpen
  \bibfield  {author} {\bibinfo {author} {\bibfnamefont {E.}~\bibnamefont
  {Berti}}, \bibinfo {author} {\bibfnamefont {V.}~\bibnamefont {Cardoso}}, \
  and\ \bibinfo {author} {\bibfnamefont {A.~O.}\ \bibnamefont {Starinets}},\
  }\href {\doibase 10.1088/0264-9381/26/16/163001} {\bibfield  {journal}
  {\bibinfo  {journal} {Class. Quant. Grav.}\ }\textbf {\bibinfo {volume}
  {26}},\ \bibinfo {pages} {163001} (\bibinfo {year} {2009})},\ \Eprint
  {http://arxiv.org/abs/0905.2975} {arXiv:0905.2975 [gr-qc]} \BibitemShut
  {NoStop}%
\bibitem [{\citenamefont {Konoplya}\ and\ \citenamefont
  {Zhidenko}(2011)}]{Konoplya:2011qq}%
  \BibitemOpen
  \bibfield  {author} {\bibinfo {author} {\bibfnamefont {R.~A.}\ \bibnamefont
  {Konoplya}}\ and\ \bibinfo {author} {\bibfnamefont {A.}~\bibnamefont
  {Zhidenko}},\ }\href {\doibase 10.1103/RevModPhys.83.793} {\bibfield
  {journal} {\bibinfo  {journal} {Rev. Mod. Phys.}\ }\textbf {\bibinfo {volume}
  {83}},\ \bibinfo {pages} {793} (\bibinfo {year} {2011})},\ \Eprint
  {http://arxiv.org/abs/1102.4014} {arXiv:1102.4014 [gr-qc]} \BibitemShut
  {NoStop}%
\bibitem [{\citenamefont {Kerr}(1963)}]{Kerr:1963ud}%
  \BibitemOpen
  \bibfield  {author} {\bibinfo {author} {\bibfnamefont {R.~P.}\ \bibnamefont
  {Kerr}},\ }\href {\doibase 10.1103/PhysRevLett.11.237} {\bibfield  {journal}
  {\bibinfo  {journal} {Phys. Rev. Lett.}\ }\textbf {\bibinfo {volume} {11}},\
  \bibinfo {pages} {237} (\bibinfo {year} {1963})}\BibitemShut {NoStop}%
\bibitem [{\citenamefont {Israel}(1967)}]{Israel:1967wq}%
  \BibitemOpen
  \bibfield  {author} {\bibinfo {author} {\bibfnamefont {W.}~\bibnamefont
  {Israel}},\ }\href {\doibase 10.1103/PhysRev.164.1776} {\bibfield  {journal}
  {\bibinfo  {journal} {Phys. Rev.}\ }\textbf {\bibinfo {volume} {164}},\
  \bibinfo {pages} {1776} (\bibinfo {year} {1967})}\BibitemShut {NoStop}%
\bibitem [{\citenamefont {Israel}(1968)}]{Israel:1967za}%
  \BibitemOpen
  \bibfield  {author} {\bibinfo {author} {\bibfnamefont {W.}~\bibnamefont
  {Israel}},\ }\href {\doibase 10.1007/BF01645859} {\bibfield  {journal}
  {\bibinfo  {journal} {Commun. Math. Phys.}\ }\textbf {\bibinfo {volume}
  {8}},\ \bibinfo {pages} {245} (\bibinfo {year} {1968})}\BibitemShut {NoStop}%
\bibitem [{\citenamefont {Carter}(1971)}]{Carter:1971zc}%
  \BibitemOpen
  \bibfield  {author} {\bibinfo {author} {\bibfnamefont {B.}~\bibnamefont
  {Carter}},\ }\href {\doibase 10.1103/PhysRevLett.26.331} {\bibfield
  {journal} {\bibinfo  {journal} {Phys. Rev. Lett.}\ }\textbf {\bibinfo
  {volume} {26}},\ \bibinfo {pages} {331} (\bibinfo {year} {1971})}\BibitemShut
  {NoStop}%
\bibitem [{\citenamefont {{Hawking}}(1972)}]{1972CMaPh..25..152H}%
  \BibitemOpen
  \bibfield  {author} {\bibinfo {author} {\bibfnamefont {S.~W.}\ \bibnamefont
  {{Hawking}}},\ }\href {\doibase 10.1007/BF01877517} {\bibfield  {journal}
  {\bibinfo  {journal} {Communications in Mathematical Physics}\ }\textbf
  {\bibinfo {volume} {25}},\ \bibinfo {pages} {152} (\bibinfo {year}
  {1972})}\BibitemShut {NoStop}%
\bibitem [{\citenamefont {Bekenstein}(1972)}]{PhysRevD.5.2403}%
  \BibitemOpen
  \bibfield  {author} {\bibinfo {author} {\bibfnamefont {J.~D.}\ \bibnamefont
  {Bekenstein}},\ }\href {\doibase 10.1103/PhysRevD.5.2403} {\bibfield
  {journal} {\bibinfo  {journal} {Phys. Rev. D}\ }\textbf {\bibinfo {volume}
  {5}},\ \bibinfo {pages} {2403} (\bibinfo {year} {1972})}\BibitemShut
  {NoStop}%
\bibitem [{\citenamefont {{Abbott}}\ \emph {et~al.}(2016)\citenamefont
  {{Abbott}}, \citenamefont {{Abbott}}, \citenamefont {{Abbott}}, \citenamefont
  {{Abernathy}}, \citenamefont {{Acernese}}, \citenamefont {{Ackley}},
  \citenamefont {{Adams}}, \citenamefont {{Adams}}, \citenamefont {{Addesso}},
  \citenamefont {{Adhikari}},\ and\ \citenamefont
  {et~al.}}]{2016PhRvL.116v1101A}%
  \BibitemOpen
  \bibfield  {author} {\bibinfo {author} {\bibfnamefont {B.~P.}\ \bibnamefont
  {{Abbott}}}, \bibinfo {author} {\bibfnamefont {R.}~\bibnamefont {{Abbott}}},
  \bibinfo {author} {\bibfnamefont {T.~D.}\ \bibnamefont {{Abbott}}}, \bibinfo
  {author} {\bibfnamefont {M.~R.}\ \bibnamefont {{Abernathy}}}, \bibinfo
  {author} {\bibfnamefont {F.}~\bibnamefont {{Acernese}}}, \bibinfo {author}
  {\bibfnamefont {K.}~\bibnamefont {{Ackley}}}, \bibinfo {author}
  {\bibfnamefont {C.}~\bibnamefont {{Adams}}}, \bibinfo {author} {\bibfnamefont
  {T.}~\bibnamefont {{Adams}}}, \bibinfo {author} {\bibfnamefont
  {P.}~\bibnamefont {{Addesso}}}, \bibinfo {author} {\bibfnamefont {R.~X.}\
  \bibnamefont {{Adhikari}}}, \ and\ \bibinfo {author} {\bibnamefont
  {et~al.}},\ }\href {\doibase 10.1103/PhysRevLett.116.221101} {\bibfield
  {journal} {\bibinfo  {journal} {Physical Review Letters}\ }\textbf {\bibinfo
  {volume} {116}},\ \bibinfo {eid} {221101} (\bibinfo {year} {2016})},\ \Eprint
  {http://arxiv.org/abs/1602.03841} {arXiv:1602.03841 [gr-qc]} \BibitemShut
  {NoStop}%
\bibitem [{\citenamefont {Bl{\'a}zquez-Salcedo}\ \emph
  {et~al.}(2016)\citenamefont {Bl{\'a}zquez-Salcedo}, \citenamefont {Macedo},
  \citenamefont {Cardoso}, \citenamefont {Ferrari}, \citenamefont {Gualtieri},
  \citenamefont {Khoo}, \citenamefont {Kunz},\ and\ \citenamefont
  {Pani}}]{Blazquez-Salcedo:2016enn}%
  \BibitemOpen
  \bibfield  {author} {\bibinfo {author} {\bibfnamefont {J.~L.}\ \bibnamefont
  {Bl{\'a}zquez-Salcedo}}, \bibinfo {author} {\bibfnamefont {C.~F.~B.}\
  \bibnamefont {Macedo}}, \bibinfo {author} {\bibfnamefont {V.}~\bibnamefont
  {Cardoso}}, \bibinfo {author} {\bibfnamefont {V.}~\bibnamefont {Ferrari}},
  \bibinfo {author} {\bibfnamefont {L.}~\bibnamefont {Gualtieri}}, \bibinfo
  {author} {\bibfnamefont {F.~S.}\ \bibnamefont {Khoo}}, \bibinfo {author}
  {\bibfnamefont {J.}~\bibnamefont {Kunz}}, \ and\ \bibinfo {author}
  {\bibfnamefont {P.}~\bibnamefont {Pani}},\ }\href {\doibase
  10.1103/PhysRevD.94.104024} {\bibfield  {journal} {\bibinfo  {journal} {Phys.
  Rev.}\ }\textbf {\bibinfo {volume} {D94}},\ \bibinfo {pages} {104024}
  (\bibinfo {year} {2016})},\ \Eprint {http://arxiv.org/abs/1609.01286}
  {arXiv:1609.01286 [gr-qc]} \BibitemShut {NoStop}%
\bibitem [{\citenamefont {Bl{\'a}zquez-Salcedo}\ \emph
  {et~al.}(2017)\citenamefont {Bl{\'a}zquez-Salcedo}, \citenamefont {Khoo},\
  and\ \citenamefont {Kunz}}]{Blazquez-Salcedo:2017txk}%
  \BibitemOpen
  \bibfield  {author} {\bibinfo {author} {\bibfnamefont {J.~L.}\ \bibnamefont
  {Bl{\'a}zquez-Salcedo}}, \bibinfo {author} {\bibfnamefont {F.~S.}\
  \bibnamefont {Khoo}}, \ and\ \bibinfo {author} {\bibfnamefont
  {J.}~\bibnamefont {Kunz}},\ }\href {\doibase 10.1103/PhysRevD.96.064008}
  {\bibfield  {journal} {\bibinfo  {journal} {Phys. Rev.}\ }\textbf {\bibinfo
  {volume} {D96}},\ \bibinfo {pages} {064008} (\bibinfo {year} {2017})},\
  \Eprint {http://arxiv.org/abs/1706.03262} {arXiv:1706.03262 [gr-qc]}
  \BibitemShut {NoStop}%
\bibitem [{\citenamefont {Silva}\ \emph {et~al.}(2018)\citenamefont {Silva},
  \citenamefont {Sakstein}, \citenamefont {Gualtieri}, \citenamefont
  {Sotiriou},\ and\ \citenamefont {Berti}}]{Silva:2017uqg}%
  \BibitemOpen
  \bibfield  {author} {\bibinfo {author} {\bibfnamefont {H.~O.}\ \bibnamefont
  {Silva}}, \bibinfo {author} {\bibfnamefont {J.}~\bibnamefont {Sakstein}},
  \bibinfo {author} {\bibfnamefont {L.}~\bibnamefont {Gualtieri}}, \bibinfo
  {author} {\bibfnamefont {T.~P.}\ \bibnamefont {Sotiriou}}, \ and\ \bibinfo
  {author} {\bibfnamefont {E.}~\bibnamefont {Berti}},\ }\href {\doibase
  10.1103/PhysRevLett.120.131104} {\bibfield  {journal} {\bibinfo  {journal}
  {Phys. Rev. Lett.}\ }\textbf {\bibinfo {volume} {120}},\ \bibinfo {pages}
  {131104} (\bibinfo {year} {2018})},\ \Eprint
  {http://arxiv.org/abs/1711.02080} {arXiv:1711.02080 [gr-qc]} \BibitemShut
  {NoStop}%
\bibitem [{\citenamefont {Antoniou}\ \emph {et~al.}(2018)\citenamefont
  {Antoniou}, \citenamefont {Bakopoulos},\ and\ \citenamefont
  {Kanti}}]{Antoniou:2017acq}%
  \BibitemOpen
  \bibfield  {author} {\bibinfo {author} {\bibfnamefont {G.}~\bibnamefont
  {Antoniou}}, \bibinfo {author} {\bibfnamefont {A.}~\bibnamefont
  {Bakopoulos}}, \ and\ \bibinfo {author} {\bibfnamefont {P.}~\bibnamefont
  {Kanti}},\ }\href {\doibase 10.1103/PhysRevLett.120.131102} {\bibfield
  {journal} {\bibinfo  {journal} {Phys. Rev. Lett.}\ }\textbf {\bibinfo
  {volume} {120}},\ \bibinfo {pages} {131102} (\bibinfo {year} {2018})},\
  \Eprint {http://arxiv.org/abs/1711.03390} {arXiv:1711.03390 [hep-th]}
  \BibitemShut {NoStop}%
\bibitem [{\citenamefont {Antoniou}\ \emph {et~al.}(2017)\citenamefont
  {Antoniou}, \citenamefont {Bakopoulos},\ and\ \citenamefont
  {Kanti}}]{Antoniou:2017hxj}%
  \BibitemOpen
  \bibfield  {author} {\bibinfo {author} {\bibfnamefont {G.}~\bibnamefont
  {Antoniou}}, \bibinfo {author} {\bibfnamefont {A.}~\bibnamefont
  {Bakopoulos}}, \ and\ \bibinfo {author} {\bibfnamefont {P.}~\bibnamefont
  {Kanti}},\ }\href@noop {} {\  (\bibinfo {year} {2017})},\ \Eprint
  {http://arxiv.org/abs/1711.07431} {arXiv:1711.07431 [hep-th]} \BibitemShut
  {NoStop}%
\bibitem [{\citenamefont {Bakopoulos}\ \emph {et~al.}(2019)\citenamefont
  {Bakopoulos}, \citenamefont {Antoniou},\ and\ \citenamefont
  {Kanti}}]{Bakopoulos:2018nui}%
  \BibitemOpen
  \bibfield  {author} {\bibinfo {author} {\bibfnamefont {A.}~\bibnamefont
  {Bakopoulos}}, \bibinfo {author} {\bibfnamefont {G.}~\bibnamefont
  {Antoniou}}, \ and\ \bibinfo {author} {\bibfnamefont {P.}~\bibnamefont
  {Kanti}},\ }\href {\doibase 10.1103/PhysRevD.99.064003} {\bibfield  {journal}
  {\bibinfo  {journal} {Phys. Rev.}\ }\textbf {\bibinfo {volume} {D99}},\
  \bibinfo {pages} {064003} (\bibinfo {year} {2019})},\ \Eprint
  {http://arxiv.org/abs/1812.06941} {arXiv:1812.06941 [hep-th]} \BibitemShut
  {NoStop}%
\bibitem [{\citenamefont {Minamitsuji}\ and\ \citenamefont
  {Ikeda}(2019)}]{Minamitsuji:2018xde}%
  \BibitemOpen
  \bibfield  {author} {\bibinfo {author} {\bibfnamefont {M.}~\bibnamefont
  {Minamitsuji}}\ and\ \bibinfo {author} {\bibfnamefont {T.}~\bibnamefont
  {Ikeda}},\ }\href {\doibase 10.1103/PhysRevD.99.044017} {\bibfield  {journal}
  {\bibinfo  {journal} {Phys. Rev.}\ }\textbf {\bibinfo {volume} {D99}},\
  \bibinfo {pages} {044017} (\bibinfo {year} {2019})},\ \Eprint
  {http://arxiv.org/abs/1812.03551} {arXiv:1812.03551 [gr-qc]} \BibitemShut
  {NoStop}%
\bibitem [{\citenamefont {Sullivan}\ \emph {et~al.}(2019)\citenamefont
  {Sullivan}, \citenamefont {Yunes},\ and\ \citenamefont
  {Sotiriou}}]{Sullivan:2019vyi}%
  \BibitemOpen
  \bibfield  {author} {\bibinfo {author} {\bibfnamefont {A.}~\bibnamefont
  {Sullivan}}, \bibinfo {author} {\bibfnamefont {N.}~\bibnamefont {Yunes}}, \
  and\ \bibinfo {author} {\bibfnamefont {T.~P.}\ \bibnamefont {Sotiriou}},\
  }\href@noop {} {\  (\bibinfo {year} {2019})},\ \Eprint
  {http://arxiv.org/abs/1903.02624} {arXiv:1903.02624 [gr-qc]} \BibitemShut
  {NoStop}%
\bibitem [{\citenamefont {Macedo}\ \emph {et~al.}(2019)\citenamefont {Macedo},
  \citenamefont {Sakstein}, \citenamefont {Berti}, \citenamefont {Gualtieri},
  \citenamefont {Silva},\ and\ \citenamefont {Sotiriou}}]{Macedo:2019sem}%
  \BibitemOpen
  \bibfield  {author} {\bibinfo {author} {\bibfnamefont {C.~F.~B.}\
  \bibnamefont {Macedo}}, \bibinfo {author} {\bibfnamefont {J.}~\bibnamefont
  {Sakstein}}, \bibinfo {author} {\bibfnamefont {E.}~\bibnamefont {Berti}},
  \bibinfo {author} {\bibfnamefont {L.}~\bibnamefont {Gualtieri}}, \bibinfo
  {author} {\bibfnamefont {H.~O.}\ \bibnamefont {Silva}}, \ and\ \bibinfo
  {author} {\bibfnamefont {T.~P.}\ \bibnamefont {Sotiriou}},\ }\href@noop {} {\
   (\bibinfo {year} {2019})},\ \Eprint {http://arxiv.org/abs/1903.06784}
  {arXiv:1903.06784 [gr-qc]} \BibitemShut {NoStop}%
\bibitem [{\citenamefont {Konoplya}(2002)}]{Konoplya:2001ji}%
  \BibitemOpen
  \bibfield  {author} {\bibinfo {author} {\bibfnamefont {R.~A.}\ \bibnamefont
  {Konoplya}},\ }\href {\doibase 10.1023/A:1015347628961} {\bibfield  {journal}
  {\bibinfo  {journal} {Gen. Rel. Grav.}\ }\textbf {\bibinfo {volume} {34}},\
  \bibinfo {pages} {329} (\bibinfo {year} {2002})},\ \Eprint
  {http://arxiv.org/abs/gr-qc/0109096} {arXiv:gr-qc/0109096 [gr-qc]}
  \BibitemShut {NoStop}%
\bibitem [{\citenamefont {Dong}\ \emph {et~al.}(2017)\citenamefont {Dong},
  \citenamefont {Sakstein},\ and\ \citenamefont {Stojkovic}}]{Dong:2017toi}%
  \BibitemOpen
  \bibfield  {author} {\bibinfo {author} {\bibfnamefont {R.}~\bibnamefont
  {Dong}}, \bibinfo {author} {\bibfnamefont {J.}~\bibnamefont {Sakstein}}, \
  and\ \bibinfo {author} {\bibfnamefont {D.}~\bibnamefont {Stojkovic}},\ }\href
  {\doibase 10.1103/PhysRevD.96.064048} {\bibfield  {journal} {\bibinfo
  {journal} {Phys. Rev.}\ }\textbf {\bibinfo {volume} {D96}},\ \bibinfo {pages}
  {064048} (\bibinfo {year} {2017})},\ \Eprint
  {http://arxiv.org/abs/1709.01641} {arXiv:1709.01641 [gr-qc]} \BibitemShut
  {NoStop}%
\bibitem [{\citenamefont {Cardoso}\ \emph {et~al.}(2018)\citenamefont
  {Cardoso}, \citenamefont {Kimura}, \citenamefont {Maselli},\ and\
  \citenamefont {Senatore}}]{Cardoso:2018ptl}%
  \BibitemOpen
  \bibfield  {author} {\bibinfo {author} {\bibfnamefont {V.}~\bibnamefont
  {Cardoso}}, \bibinfo {author} {\bibfnamefont {M.}~\bibnamefont {Kimura}},
  \bibinfo {author} {\bibfnamefont {A.}~\bibnamefont {Maselli}}, \ and\
  \bibinfo {author} {\bibfnamefont {L.}~\bibnamefont {Senatore}},\ }\href
  {\doibase 10.1103/PhysRevLett.121.251105} {\bibfield  {journal} {\bibinfo
  {journal} {Phys. Rev. Lett.}\ }\textbf {\bibinfo {volume} {121}},\ \bibinfo
  {pages} {251105} (\bibinfo {year} {2018})},\ \Eprint
  {http://arxiv.org/abs/1808.08962} {arXiv:1808.08962 [gr-qc]} \BibitemShut
  {NoStop}%
\bibitem [{\citenamefont {Brito}\ and\ \citenamefont
  {Pacilio}(2018)}]{Brito:2018hjh}%
  \BibitemOpen
  \bibfield  {author} {\bibinfo {author} {\bibfnamefont {R.}~\bibnamefont
  {Brito}}\ and\ \bibinfo {author} {\bibfnamefont {C.}~\bibnamefont
  {Pacilio}},\ }\href {\doibase 10.1103/PhysRevD.98.104042} {\bibfield
  {journal} {\bibinfo  {journal} {Phys. Rev.}\ }\textbf {\bibinfo {volume}
  {D98}},\ \bibinfo {pages} {104042} (\bibinfo {year} {2018})},\ \Eprint
  {http://arxiv.org/abs/1807.09081} {arXiv:1807.09081 [gr-qc]} \BibitemShut
  {NoStop}%
\bibitem [{\citenamefont {Franciolini}\ \emph {et~al.}(2019)\citenamefont
  {Franciolini}, \citenamefont {Hui}, \citenamefont {Penco}, \citenamefont
  {Santoni},\ and\ \citenamefont {Trincherini}}]{Franciolini:2018uyq}%
  \BibitemOpen
  \bibfield  {author} {\bibinfo {author} {\bibfnamefont {G.}~\bibnamefont
  {Franciolini}}, \bibinfo {author} {\bibfnamefont {L.}~\bibnamefont {Hui}},
  \bibinfo {author} {\bibfnamefont {R.}~\bibnamefont {Penco}}, \bibinfo
  {author} {\bibfnamefont {L.}~\bibnamefont {Santoni}}, \ and\ \bibinfo
  {author} {\bibfnamefont {E.}~\bibnamefont {Trincherini}},\ }\href {\doibase
  10.1007/JHEP02(2019)127} {\bibfield  {journal} {\bibinfo  {journal} {JHEP}\
  }\textbf {\bibinfo {volume} {02}},\ \bibinfo {pages} {127} (\bibinfo {year}
  {2019})},\ \Eprint {http://arxiv.org/abs/1810.07706} {arXiv:1810.07706
  [hep-th]} \BibitemShut {NoStop}%
\bibitem [{\citenamefont {Barausse}\ and\ \citenamefont
  {Sotiriou}(2008)}]{Barausse:2008xv}%
  \BibitemOpen
  \bibfield  {author} {\bibinfo {author} {\bibfnamefont {E.}~\bibnamefont
  {Barausse}}\ and\ \bibinfo {author} {\bibfnamefont {T.~P.}\ \bibnamefont
  {Sotiriou}},\ }\href {\doibase 10.1103/PhysRevLett.101.099001} {\bibfield
  {journal} {\bibinfo  {journal} {Phys. Rev. Lett.}\ }\textbf {\bibinfo
  {volume} {101}},\ \bibinfo {pages} {099001} (\bibinfo {year} {2008})},\
  \Eprint {http://arxiv.org/abs/0803.3433} {arXiv:0803.3433 [gr-qc]}
  \BibitemShut {NoStop}%
\bibitem [{\citenamefont {Molina}\ \emph {et~al.}(2010)\citenamefont {Molina},
  \citenamefont {Pani}, \citenamefont {Cardoso},\ and\ \citenamefont
  {Gualtieri}}]{Molina:2010fb}%
  \BibitemOpen
  \bibfield  {author} {\bibinfo {author} {\bibfnamefont {C.}~\bibnamefont
  {Molina}}, \bibinfo {author} {\bibfnamefont {P.}~\bibnamefont {Pani}},
  \bibinfo {author} {\bibfnamefont {V.}~\bibnamefont {Cardoso}}, \ and\
  \bibinfo {author} {\bibfnamefont {L.}~\bibnamefont {Gualtieri}},\ }\href
  {\doibase 10.1103/PhysRevD.81.124021} {\bibfield  {journal} {\bibinfo
  {journal} {Phys. Rev.}\ }\textbf {\bibinfo {volume} {D81}},\ \bibinfo {pages}
  {124021} (\bibinfo {year} {2010})},\ \Eprint {http://arxiv.org/abs/1004.4007}
  {arXiv:1004.4007 [gr-qc]} \BibitemShut {NoStop}%
\bibitem [{\citenamefont {Tattersall}\ \emph
  {et~al.}(2018{\natexlab{a}})\citenamefont {Tattersall}, \citenamefont
  {Ferreira},\ and\ \citenamefont {Lagos}}]{Tattersall:2017erk}%
  \BibitemOpen
  \bibfield  {author} {\bibinfo {author} {\bibfnamefont {O.~J.}\ \bibnamefont
  {Tattersall}}, \bibinfo {author} {\bibfnamefont {P.~G.}\ \bibnamefont
  {Ferreira}}, \ and\ \bibinfo {author} {\bibfnamefont {M.}~\bibnamefont
  {Lagos}},\ }\href {\doibase 10.1103/PhysRevD.97.044021} {\bibfield  {journal}
  {\bibinfo  {journal} {Phys. Rev.}\ }\textbf {\bibinfo {volume} {D97}},\
  \bibinfo {pages} {044021} (\bibinfo {year} {2018}{\natexlab{a}})},\ \Eprint
  {http://arxiv.org/abs/1711.01992} {arXiv:1711.01992 [gr-qc]} \BibitemShut
  {NoStop}%
\bibitem [{\citenamefont {Tattersall}\ and\ \citenamefont
  {Ferreira}(2018)}]{Tattersall:2018nve}%
  \BibitemOpen
  \bibfield  {author} {\bibinfo {author} {\bibfnamefont {O.~J.}\ \bibnamefont
  {Tattersall}}\ and\ \bibinfo {author} {\bibfnamefont {P.~G.}\ \bibnamefont
  {Ferreira}},\ }\href {\doibase 10.1103/PhysRevD.97.104047} {\bibfield
  {journal} {\bibinfo  {journal} {Phys. Rev.}\ }\textbf {\bibinfo {volume}
  {D97}},\ \bibinfo {pages} {104047} (\bibinfo {year} {2018})},\ \Eprint
  {http://arxiv.org/abs/1804.08950} {arXiv:1804.08950 [gr-qc]} \BibitemShut
  {NoStop}%
\bibitem [{\citenamefont {Tattersall}\ and\ \citenamefont
  {Ferreira}(2019)}]{Tattersall:2019pvx}%
  \BibitemOpen
  \bibfield  {author} {\bibinfo {author} {\bibfnamefont {O.~J.}\ \bibnamefont
  {Tattersall}}\ and\ \bibinfo {author} {\bibfnamefont {P.~G.}\ \bibnamefont
  {Ferreira}},\ }\href {\doibase 10.1103/PhysRevD.99.104082} {\bibfield
  {journal} {\bibinfo  {journal} {Phys. Rev.}\ }\textbf {\bibinfo {volume}
  {D99}},\ \bibinfo {pages} {104082} (\bibinfo {year} {2019})},\ \Eprint
  {http://arxiv.org/abs/1904.05112} {arXiv:1904.05112 [gr-qc]} \BibitemShut
  {NoStop}%
\bibitem [{\citenamefont {Isi}\ \emph {et~al.}(2019)\citenamefont {Isi},
  \citenamefont {Giesler}, \citenamefont {Farr}, \citenamefont {Scheel},\ and\
  \citenamefont {Teukolsky}}]{Isi:2019aib}%
  \BibitemOpen
  \bibfield  {author} {\bibinfo {author} {\bibfnamefont {M.}~\bibnamefont
  {Isi}}, \bibinfo {author} {\bibfnamefont {M.}~\bibnamefont {Giesler}},
  \bibinfo {author} {\bibfnamefont {W.~M.}\ \bibnamefont {Farr}}, \bibinfo
  {author} {\bibfnamefont {M.~A.}\ \bibnamefont {Scheel}}, \ and\ \bibinfo
  {author} {\bibfnamefont {S.~A.}\ \bibnamefont {Teukolsky}},\ }\href@noop {}
  {\  (\bibinfo {year} {2019})},\ \Eprint {http://arxiv.org/abs/1905.00869}
  {arXiv:1905.00869 [gr-qc]} \BibitemShut {NoStop}%
\bibitem [{\citenamefont {Kojima}(1992)}]{Kojima:1992ie}%
  \BibitemOpen
  \bibfield  {author} {\bibinfo {author} {\bibfnamefont {Y.}~\bibnamefont
  {Kojima}},\ }\href {\doibase 10.1103/PhysRevD.46.4289} {\bibfield  {journal}
  {\bibinfo  {journal} {Phys. Rev.}\ }\textbf {\bibinfo {volume} {D46}},\
  \bibinfo {pages} {4289} (\bibinfo {year} {1992})}\BibitemShut {NoStop}%
\bibitem [{\citenamefont {Pani}(2012)}]{Pani:2012zz}%
  \BibitemOpen
  \bibfield  {author} {\bibinfo {author} {\bibfnamefont {P.}~\bibnamefont
  {Pani}},\ }\href {\doibase 10.1140/epjp/i2012-12067-1} {\bibfield  {journal}
  {\bibinfo  {journal} {Eur. Phys. J. Plus}\ }\textbf {\bibinfo {volume}
  {127}},\ \bibinfo {pages} {67} (\bibinfo {year} {2012})}\BibitemShut
  {NoStop}%
\bibitem [{\citenamefont {Pani}\ \emph {et~al.}(2012)\citenamefont {Pani},
  \citenamefont {Cardoso}, \citenamefont {Gualtieri}, \citenamefont {Berti},\
  and\ \citenamefont {Ishibashi}}]{Pani:2012bp}%
  \BibitemOpen
  \bibfield  {author} {\bibinfo {author} {\bibfnamefont {P.}~\bibnamefont
  {Pani}}, \bibinfo {author} {\bibfnamefont {V.}~\bibnamefont {Cardoso}},
  \bibinfo {author} {\bibfnamefont {L.}~\bibnamefont {Gualtieri}}, \bibinfo
  {author} {\bibfnamefont {E.}~\bibnamefont {Berti}}, \ and\ \bibinfo {author}
  {\bibfnamefont {A.}~\bibnamefont {Ishibashi}},\ }\href {\doibase
  10.1103/PhysRevD.86.104017} {\bibfield  {journal} {\bibinfo  {journal} {Phys.
  Rev.}\ }\textbf {\bibinfo {volume} {D86}},\ \bibinfo {pages} {104017}
  (\bibinfo {year} {2012})},\ \Eprint {http://arxiv.org/abs/1209.0773}
  {arXiv:1209.0773 [gr-qc]} \BibitemShut {NoStop}%
\bibitem [{\citenamefont {Tattersall}(2018)}]{Tattersall:2018axd}%
  \BibitemOpen
  \bibfield  {author} {\bibinfo {author} {\bibfnamefont {O.~J.}\ \bibnamefont
  {Tattersall}},\ }\href {\doibase 10.1103/PhysRevD.98.104013} {\bibfield
  {journal} {\bibinfo  {journal} {Phys. Rev.}\ }\textbf {\bibinfo {volume}
  {D98}},\ \bibinfo {pages} {104013} (\bibinfo {year} {2018})},\ \Eprint
  {http://arxiv.org/abs/1808.10758} {arXiv:1808.10758 [gr-qc]} \BibitemShut
  {NoStop}%
\bibitem [{\citenamefont {Horndeski}(1974)}]{Horndeski:1974wa}%
  \BibitemOpen
  \bibfield  {author} {\bibinfo {author} {\bibfnamefont {G.~W.}\ \bibnamefont
  {Horndeski}},\ }\href {\doibase 10.1007/BF01807638} {\bibfield  {journal}
  {\bibinfo  {journal} {Int. J. Theor. Phys.}\ }\textbf {\bibinfo {volume}
  {10}},\ \bibinfo {pages} {363} (\bibinfo {year} {1974})}\BibitemShut
  {NoStop}%
\bibitem [{\citenamefont {Kobayashi}(2019)}]{Kobayashi:2019hrl}%
  \BibitemOpen
  \bibfield  {author} {\bibinfo {author} {\bibfnamefont {T.}~\bibnamefont
  {Kobayashi}},\ }\href {\doibase 10.1088/1361-6633/ab2429} {\bibfield
  {journal} {\bibinfo  {journal} {Rept. Prog. Phys.}\ }\textbf {\bibinfo
  {volume} {82}},\ \bibinfo {pages} {086901} (\bibinfo {year} {2019})},\
  \Eprint {http://arxiv.org/abs/1901.07183} {arXiv:1901.07183 [gr-qc]}
  \BibitemShut {NoStop}%
\bibitem [{\citenamefont {Cardoso}\ \emph {et~al.}(2019)\citenamefont
  {Cardoso}, \citenamefont {Kimura}, \citenamefont {Maselli}, \citenamefont
  {Berti}, \citenamefont {Macedo},\ and\ \citenamefont
  {McManus}}]{Cardoso:2019mqo}%
  \BibitemOpen
  \bibfield  {author} {\bibinfo {author} {\bibfnamefont {V.}~\bibnamefont
  {Cardoso}}, \bibinfo {author} {\bibfnamefont {M.}~\bibnamefont {Kimura}},
  \bibinfo {author} {\bibfnamefont {A.}~\bibnamefont {Maselli}}, \bibinfo
  {author} {\bibfnamefont {E.}~\bibnamefont {Berti}}, \bibinfo {author}
  {\bibfnamefont {C.~F.~B.}\ \bibnamefont {Macedo}}, \ and\ \bibinfo {author}
  {\bibfnamefont {R.}~\bibnamefont {McManus}},\ }\href@noop {} {\  (\bibinfo
  {year} {2019})},\ \Eprint {http://arxiv.org/abs/1901.01265} {arXiv:1901.01265
  [gr-qc]} \BibitemShut {NoStop}%
\bibitem [{\citenamefont {Kobayashi}\ \emph {et~al.}(2011)\citenamefont
  {Kobayashi}, \citenamefont {Yamaguchi},\ and\ \citenamefont
  {Yokoyama}}]{Kobayashi:2011nu}%
  \BibitemOpen
  \bibfield  {author} {\bibinfo {author} {\bibfnamefont {T.}~\bibnamefont
  {Kobayashi}}, \bibinfo {author} {\bibfnamefont {M.}~\bibnamefont
  {Yamaguchi}}, \ and\ \bibinfo {author} {\bibfnamefont {J.}~\bibnamefont
  {Yokoyama}},\ }\href {\doibase 10.1143/PTP.126.511} {\bibfield  {journal}
  {\bibinfo  {journal} {Prog. Theor. Phys.}\ }\textbf {\bibinfo {volume}
  {126}},\ \bibinfo {pages} {511} (\bibinfo {year} {2011})},\ \Eprint
  {http://arxiv.org/abs/1105.5723} {arXiv:1105.5723 [hep-th]} \BibitemShut
  {NoStop}%
\bibitem [{\citenamefont {Langlois}\ and\ \citenamefont
  {Noui}(2016)}]{Langlois:2015cwa}%
  \BibitemOpen
  \bibfield  {author} {\bibinfo {author} {\bibfnamefont {D.}~\bibnamefont
  {Langlois}}\ and\ \bibinfo {author} {\bibfnamefont {K.}~\bibnamefont
  {Noui}},\ }\href {\doibase 10.1088/1475-7516/2016/02/034} {\bibfield
  {journal} {\bibinfo  {journal} {JCAP}\ }\textbf {\bibinfo {volume} {1602}},\
  \bibinfo {pages} {034} (\bibinfo {year} {2016})},\ \Eprint
  {http://arxiv.org/abs/1510.06930} {arXiv:1510.06930 [gr-qc]} \BibitemShut
  {NoStop}%
\bibitem [{\citenamefont {Zumalacarregui}\ and\ \citenamefont
  {Garcia-Bellido}(2014)}]{Zumalacarregui:2013pma}%
  \BibitemOpen
  \bibfield  {author} {\bibinfo {author} {\bibfnamefont {M.}~\bibnamefont
  {Zumalacarregui}}\ and\ \bibinfo {author} {\bibfnamefont {J.}~\bibnamefont
  {Garcia-Bellido}},\ }\href {\doibase 10.1103/PhysRevD.89.064046} {\bibfield
  {journal} {\bibinfo  {journal} {Phys. Rev.}\ }\textbf {\bibinfo {volume}
  {D89}},\ \bibinfo {pages} {064046} (\bibinfo {year} {2014})},\ \Eprint
  {http://arxiv.org/abs/1308.4685} {arXiv:1308.4685 [gr-qc]} \BibitemShut
  {NoStop}%
\bibitem [{\citenamefont {Gleyzes}\ \emph
  {et~al.}(2015{\natexlab{a}})\citenamefont {Gleyzes}, \citenamefont
  {Langlois}, \citenamefont {Piazza},\ and\ \citenamefont
  {Vernizzi}}]{Gleyzes:2014qga}%
  \BibitemOpen
  \bibfield  {author} {\bibinfo {author} {\bibfnamefont {J.}~\bibnamefont
  {Gleyzes}}, \bibinfo {author} {\bibfnamefont {D.}~\bibnamefont {Langlois}},
  \bibinfo {author} {\bibfnamefont {F.}~\bibnamefont {Piazza}}, \ and\ \bibinfo
  {author} {\bibfnamefont {F.}~\bibnamefont {Vernizzi}},\ }\href {\doibase
  10.1088/1475-7516/2015/02/018} {\bibfield  {journal} {\bibinfo  {journal}
  {JCAP}\ }\textbf {\bibinfo {volume} {1502}},\ \bibinfo {pages} {018}
  (\bibinfo {year} {2015}{\natexlab{a}})},\ \Eprint
  {http://arxiv.org/abs/1408.1952} {arXiv:1408.1952 [astro-ph.CO]} \BibitemShut
  {NoStop}%
\bibitem [{\citenamefont {Gleyzes}\ \emph
  {et~al.}(2015{\natexlab{b}})\citenamefont {Gleyzes}, \citenamefont
  {Langlois}, \citenamefont {Piazza},\ and\ \citenamefont
  {Vernizzi}}]{Gleyzes:2014dya}%
  \BibitemOpen
  \bibfield  {author} {\bibinfo {author} {\bibfnamefont {J.}~\bibnamefont
  {Gleyzes}}, \bibinfo {author} {\bibfnamefont {D.}~\bibnamefont {Langlois}},
  \bibinfo {author} {\bibfnamefont {F.}~\bibnamefont {Piazza}}, \ and\ \bibinfo
  {author} {\bibfnamefont {F.}~\bibnamefont {Vernizzi}},\ }\href {\doibase
  10.1103/PhysRevLett.114.211101} {\bibfield  {journal} {\bibinfo  {journal}
  {Phys. Rev. Lett.}\ }\textbf {\bibinfo {volume} {114}},\ \bibinfo {pages}
  {211101} (\bibinfo {year} {2015}{\natexlab{b}})},\ \Eprint
  {http://arxiv.org/abs/1404.6495} {arXiv:1404.6495 [hep-th]} \BibitemShut
  {NoStop}%
\bibitem [{\citenamefont {Ben~Achour}\ \emph {et~al.}(2016)\citenamefont
  {Ben~Achour}, \citenamefont {Langlois},\ and\ \citenamefont
  {Noui}}]{Achour:2016rkg}%
  \BibitemOpen
  \bibfield  {author} {\bibinfo {author} {\bibfnamefont {J.}~\bibnamefont
  {Ben~Achour}}, \bibinfo {author} {\bibfnamefont {D.}~\bibnamefont
  {Langlois}}, \ and\ \bibinfo {author} {\bibfnamefont {K.}~\bibnamefont
  {Noui}},\ }\href {\doibase 10.1103/PhysRevD.93.124005} {\bibfield  {journal}
  {\bibinfo  {journal} {Phys. Rev.}\ }\textbf {\bibinfo {volume} {D93}},\
  \bibinfo {pages} {124005} (\bibinfo {year} {2016})},\ \Eprint
  {http://arxiv.org/abs/1602.08398} {arXiv:1602.08398 [gr-qc]} \BibitemShut
  {NoStop}%
\bibitem [{\citenamefont {Kase}\ and\ \citenamefont
  {Tsujikawa}(2013)}]{Kase:2013uja}%
  \BibitemOpen
  \bibfield  {author} {\bibinfo {author} {\bibfnamefont {R.}~\bibnamefont
  {Kase}}\ and\ \bibinfo {author} {\bibfnamefont {S.}~\bibnamefont
  {Tsujikawa}},\ }\href {\doibase 10.1088/1475-7516/2013/08/054} {\bibfield
  {journal} {\bibinfo  {journal} {JCAP}\ }\textbf {\bibinfo {volume} {1308}},\
  \bibinfo {pages} {054} (\bibinfo {year} {2013})},\ \Eprint
  {http://arxiv.org/abs/1306.6401} {arXiv:1306.6401 [gr-qc]} \BibitemShut
  {NoStop}%
\bibitem [{\citenamefont {McManus}\ \emph {et~al.}(2016)\citenamefont
  {McManus}, \citenamefont {Lombriser},\ and\ \citenamefont
  {Penarrubia}}]{McManus:2016kxu}%
  \BibitemOpen
  \bibfield  {author} {\bibinfo {author} {\bibfnamefont {R.}~\bibnamefont
  {McManus}}, \bibinfo {author} {\bibfnamefont {L.}~\bibnamefont {Lombriser}},
  \ and\ \bibinfo {author} {\bibfnamefont {J.}~\bibnamefont {Penarrubia}},\
  }\href {\doibase 10.1088/1475-7516/2016/11/006} {\bibfield  {journal}
  {\bibinfo  {journal} {JCAP}\ }\textbf {\bibinfo {volume} {1611}},\ \bibinfo
  {pages} {006} (\bibinfo {year} {2016})},\ \Eprint
  {http://arxiv.org/abs/1606.03282} {arXiv:1606.03282 [gr-qc]} \BibitemShut
  {NoStop}%
\bibitem [{\citenamefont {et.
  al.}(2017{\natexlab{a}})}]{PhysRevLett.119.161101}%
  \BibitemOpen
  \bibfield  {author} {\bibinfo {author} {\bibfnamefont {B.~P.~A.}\
  \bibnamefont {et. al.}} (\bibinfo {collaboration} {LIGO Scientific
  Collaboration and Virgo Collaboration}),\ }\href {\doibase
  10.1103/PhysRevLett.119.161101} {\bibfield  {journal} {\bibinfo  {journal}
  {Phys. Rev. Lett.}\ }\textbf {\bibinfo {volume} {119}},\ \bibinfo {pages}
  {161101} (\bibinfo {year} {2017}{\natexlab{a}})}\BibitemShut {NoStop}%
\bibitem [{\citenamefont {et. al.}(2017{\natexlab{b}})}]{2041-8205-848-2-L12}%
  \BibitemOpen
  \bibfield  {author} {\bibinfo {author} {\bibfnamefont {B.~P.~A.}\
  \bibnamefont {et. al.}},\ }\href
  {http://stacks.iop.org/2041-8205/848/i=2/a=L12} {\bibfield  {journal}
  {\bibinfo  {journal} {The Astrophysical Journal Letters}\ }\textbf {\bibinfo
  {volume} {848}},\ \bibinfo {pages} {L12} (\bibinfo {year}
  {2017}{\natexlab{b}})}\BibitemShut {NoStop}%
\bibitem [{\citenamefont {et. al.}(2017{\natexlab{c}})}]{2041-8205-848-2-L13}%
  \BibitemOpen
  \bibfield  {author} {\bibinfo {author} {\bibfnamefont {B.~P.~A.}\
  \bibnamefont {et. al.}},\ }\href
  {http://stacks.iop.org/2041-8205/848/i=2/a=L13} {\bibfield  {journal}
  {\bibinfo  {journal} {The Astrophysical Journal Letters}\ }\textbf {\bibinfo
  {volume} {848}},\ \bibinfo {pages} {L13} (\bibinfo {year}
  {2017}{\natexlab{c}})}\BibitemShut {NoStop}%
\bibitem [{\citenamefont {et. al.}(2017{\natexlab{d}})}]{2041-8205-848-2-L14}%
  \BibitemOpen
  \bibfield  {author} {\bibinfo {author} {\bibfnamefont {A.~G.}\ \bibnamefont
  {et. al.}},\ }\href {http://stacks.iop.org/2041-8205/848/i=2/a=L14}
  {\bibfield  {journal} {\bibinfo  {journal} {The Astrophysical Journal
  Letters}\ }\textbf {\bibinfo {volume} {848}},\ \bibinfo {pages} {L14}
  (\bibinfo {year} {2017}{\natexlab{d}})}\BibitemShut {NoStop}%
\bibitem [{\citenamefont {et. al.}(2017{\natexlab{e}})}]{2041-8205-848-2-L15}%
  \BibitemOpen
  \bibfield  {author} {\bibinfo {author} {\bibfnamefont {V.~S.}\ \bibnamefont
  {et. al.}},\ }\href {http://stacks.iop.org/2041-8205/848/i=2/a=L15}
  {\bibfield  {journal} {\bibinfo  {journal} {The Astrophysical Journal
  Letters}\ }\textbf {\bibinfo {volume} {848}},\ \bibinfo {pages} {L15}
  (\bibinfo {year} {2017}{\natexlab{e}})}\BibitemShut {NoStop}%
\bibitem [{\citenamefont {{Coulter}}\ \emph {et~al.}(2017)\citenamefont
  {{Coulter}}, \citenamefont {{Foley}}, \citenamefont {{Kilpatrick}},
  \citenamefont {{Drout}}, \citenamefont {{Piro}}, \citenamefont {{Shappee}},
  \citenamefont {{Siebert}}, \citenamefont {{Simon}}, \citenamefont {{Ulloa}},
  \citenamefont {{Kasen}}, \citenamefont {{Madore}}, \citenamefont
  {{Murguia-Berthier}}, \citenamefont {{Pan}}, \citenamefont {{Prochaska}},
  \citenamefont {{Ramirez-Ruiz}}, \citenamefont {{Rest}},\ and\ \citenamefont
  {{Rojas-Bravo}}}]{2017Sci...358.1556C}%
  \BibitemOpen
  \bibfield  {author} {\bibinfo {author} {\bibfnamefont {D.~A.}\ \bibnamefont
  {{Coulter}}}, \bibinfo {author} {\bibfnamefont {R.~J.}\ \bibnamefont
  {{Foley}}}, \bibinfo {author} {\bibfnamefont {C.~D.}\ \bibnamefont
  {{Kilpatrick}}}, \bibinfo {author} {\bibfnamefont {M.~R.}\ \bibnamefont
  {{Drout}}}, \bibinfo {author} {\bibfnamefont {A.~L.}\ \bibnamefont {{Piro}}},
  \bibinfo {author} {\bibfnamefont {B.~J.}\ \bibnamefont {{Shappee}}}, \bibinfo
  {author} {\bibfnamefont {M.~R.}\ \bibnamefont {{Siebert}}}, \bibinfo {author}
  {\bibfnamefont {J.~D.}\ \bibnamefont {{Simon}}}, \bibinfo {author}
  {\bibfnamefont {N.}~\bibnamefont {{Ulloa}}}, \bibinfo {author} {\bibfnamefont
  {D.}~\bibnamefont {{Kasen}}}, \bibinfo {author} {\bibfnamefont {B.~F.}\
  \bibnamefont {{Madore}}}, \bibinfo {author} {\bibfnamefont {A.}~\bibnamefont
  {{Murguia-Berthier}}}, \bibinfo {author} {\bibfnamefont {Y.-C.}\ \bibnamefont
  {{Pan}}}, \bibinfo {author} {\bibfnamefont {J.~X.}\ \bibnamefont
  {{Prochaska}}}, \bibinfo {author} {\bibfnamefont {E.}~\bibnamefont
  {{Ramirez-Ruiz}}}, \bibinfo {author} {\bibfnamefont {A.}~\bibnamefont
  {{Rest}}}, \ and\ \bibinfo {author} {\bibfnamefont {C.}~\bibnamefont
  {{Rojas-Bravo}}},\ }\href {\doibase 10.1126/science.aap9811} {\bibfield
  {journal} {\bibinfo  {journal} {Science}\ }\textbf {\bibinfo {volume}
  {358}},\ \bibinfo {pages} {1556} (\bibinfo {year} {2017})},\ \Eprint
  {http://arxiv.org/abs/1710.05452} {arXiv:1710.05452 [astro-ph.HE]}
  \BibitemShut {NoStop}%
\bibitem [{\citenamefont {Lombriser}\ and\ \citenamefont
  {Taylor}(2016)}]{Lombriser:2015sxa}%
  \BibitemOpen
  \bibfield  {author} {\bibinfo {author} {\bibfnamefont {L.}~\bibnamefont
  {Lombriser}}\ and\ \bibinfo {author} {\bibfnamefont {A.}~\bibnamefont
  {Taylor}},\ }\href {\doibase 10.1088/1475-7516/2016/03/031} {\bibfield
  {journal} {\bibinfo  {journal} {JCAP}\ }\textbf {\bibinfo {volume} {1603}},\
  \bibinfo {pages} {031} (\bibinfo {year} {2016})},\ \Eprint
  {http://arxiv.org/abs/1509.08458} {arXiv:1509.08458 [astro-ph.CO]}
  \BibitemShut {NoStop}%
\bibitem [{\citenamefont {Lombriser}\ and\ \citenamefont
  {Lima}(2017)}]{Lombriser:2016yzn}%
  \BibitemOpen
  \bibfield  {author} {\bibinfo {author} {\bibfnamefont {L.}~\bibnamefont
  {Lombriser}}\ and\ \bibinfo {author} {\bibfnamefont {N.~A.}\ \bibnamefont
  {Lima}},\ }\href {\doibase 10.1016/j.physletb.2016.12.048} {\bibfield
  {journal} {\bibinfo  {journal} {Phys. Lett.}\ }\textbf {\bibinfo {volume}
  {B765}},\ \bibinfo {pages} {382} (\bibinfo {year} {2017})},\ \Eprint
  {http://arxiv.org/abs/1602.07670} {arXiv:1602.07670 [astro-ph.CO]}
  \BibitemShut {NoStop}%
\bibitem [{\citenamefont {Baker}\ \emph {et~al.}(2017)\citenamefont {Baker},
  \citenamefont {Bellini}, \citenamefont {Ferreira}, \citenamefont {Lagos},
  \citenamefont {Noller},\ and\ \citenamefont {Sawicki}}]{Baker:2017hug}%
  \BibitemOpen
  \bibfield  {author} {\bibinfo {author} {\bibfnamefont {T.}~\bibnamefont
  {Baker}}, \bibinfo {author} {\bibfnamefont {E.}~\bibnamefont {Bellini}},
  \bibinfo {author} {\bibfnamefont {P.~G.}\ \bibnamefont {Ferreira}}, \bibinfo
  {author} {\bibfnamefont {M.}~\bibnamefont {Lagos}}, \bibinfo {author}
  {\bibfnamefont {J.}~\bibnamefont {Noller}}, \ and\ \bibinfo {author}
  {\bibfnamefont {I.}~\bibnamefont {Sawicki}},\ }\href@noop {} {\  (\bibinfo
  {year} {2017})},\ \Eprint {http://arxiv.org/abs/1710.06394} {arXiv:1710.06394
  [astro-ph.CO]} \BibitemShut {NoStop}%
\bibitem [{\citenamefont {Creminelli}\ and\ \citenamefont
  {Vernizzi}(2017)}]{Creminelli:2017sry}%
  \BibitemOpen
  \bibfield  {author} {\bibinfo {author} {\bibfnamefont {P.}~\bibnamefont
  {Creminelli}}\ and\ \bibinfo {author} {\bibfnamefont {F.}~\bibnamefont
  {Vernizzi}},\ }\href {\doibase 10.1103/PhysRevLett.119.251302} {\bibfield
  {journal} {\bibinfo  {journal} {Phys. Rev. Lett.}\ }\textbf {\bibinfo
  {volume} {119}},\ \bibinfo {pages} {251302} (\bibinfo {year} {2017})},\
  \Eprint {http://arxiv.org/abs/1710.05877} {arXiv:1710.05877 [astro-ph.CO]}
  \BibitemShut {NoStop}%
\bibitem [{\citenamefont {Sakstein}\ and\ \citenamefont
  {Jain}(2017)}]{Sakstein:2017xjx}%
  \BibitemOpen
  \bibfield  {author} {\bibinfo {author} {\bibfnamefont {J.}~\bibnamefont
  {Sakstein}}\ and\ \bibinfo {author} {\bibfnamefont {B.}~\bibnamefont
  {Jain}},\ }\href {\doibase 10.1103/PhysRevLett.119.251303} {\bibfield
  {journal} {\bibinfo  {journal} {Phys. Rev. Lett.}\ }\textbf {\bibinfo
  {volume} {119}},\ \bibinfo {pages} {251303} (\bibinfo {year} {2017})},\
  \Eprint {http://arxiv.org/abs/1710.05893} {arXiv:1710.05893 [astro-ph.CO]}
  \BibitemShut {NoStop}%
\bibitem [{\citenamefont {Ezquiaga}\ and\ \citenamefont
  {Zumalacarregui}(2017)}]{Ezquiaga:2017ekz}%
  \BibitemOpen
  \bibfield  {author} {\bibinfo {author} {\bibfnamefont {J.~M.}\ \bibnamefont
  {Ezquiaga}}\ and\ \bibinfo {author} {\bibfnamefont {M.}~\bibnamefont
  {Zumalacarregui}},\ }\href {\doibase 10.1103/PhysRevLett.119.251304}
  {\bibfield  {journal} {\bibinfo  {journal} {Phys. Rev. Lett.}\ }\textbf
  {\bibinfo {volume} {119}},\ \bibinfo {pages} {251304} (\bibinfo {year}
  {2017})},\ \Eprint {http://arxiv.org/abs/1710.05901} {arXiv:1710.05901
  [astro-ph.CO]} \BibitemShut {NoStop}%
\bibitem [{\citenamefont {de~Rham}\ and\ \citenamefont
  {Melville}(2018)}]{deRham:2018red}%
  \BibitemOpen
  \bibfield  {author} {\bibinfo {author} {\bibfnamefont {C.}~\bibnamefont
  {de~Rham}}\ and\ \bibinfo {author} {\bibfnamefont {S.}~\bibnamefont
  {Melville}},\ }\href {\doibase 10.1103/PhysRevLett.121.221101} {\bibfield
  {journal} {\bibinfo  {journal} {Phys. Rev. Lett.}\ }\textbf {\bibinfo
  {volume} {121}},\ \bibinfo {pages} {221101} (\bibinfo {year} {2018})},\
  \Eprint {http://arxiv.org/abs/1806.09417} {arXiv:1806.09417 [hep-th]}
  \BibitemShut {NoStop}%
\bibitem [{\citenamefont {Tattersall}\ \emph
  {et~al.}(2018{\natexlab{b}})\citenamefont {Tattersall}, \citenamefont
  {Ferreira},\ and\ \citenamefont {Lagos}}]{PhysRevD.97.084005}%
  \BibitemOpen
  \bibfield  {author} {\bibinfo {author} {\bibfnamefont {O.~J.}\ \bibnamefont
  {Tattersall}}, \bibinfo {author} {\bibfnamefont {P.~G.}\ \bibnamefont
  {Ferreira}}, \ and\ \bibinfo {author} {\bibfnamefont {M.}~\bibnamefont
  {Lagos}},\ }\href {\doibase 10.1103/PhysRevD.97.084005} {\bibfield  {journal}
  {\bibinfo  {journal} {Phys. Rev. D}\ }\textbf {\bibinfo {volume} {97}},\
  \bibinfo {pages} {084005} (\bibinfo {year} {2018}{\natexlab{b}})}\BibitemShut
  {NoStop}%
\bibitem [{\citenamefont {Kobayashi}\ \emph {et~al.}(2012)\citenamefont
  {Kobayashi}, \citenamefont {Motohashi},\ and\ \citenamefont
  {Suyama}}]{Kobayashi:2012kh}%
  \BibitemOpen
  \bibfield  {author} {\bibinfo {author} {\bibfnamefont {T.}~\bibnamefont
  {Kobayashi}}, \bibinfo {author} {\bibfnamefont {H.}~\bibnamefont
  {Motohashi}}, \ and\ \bibinfo {author} {\bibfnamefont {T.}~\bibnamefont
  {Suyama}},\ }\href {\doibase 10.1103/PhysRevD.85.084025} {\bibfield
  {journal} {\bibinfo  {journal} {Phys. Rev.}\ }\textbf {\bibinfo {volume}
  {D85}},\ \bibinfo {pages} {084025} (\bibinfo {year} {2012})},\ \Eprint
  {http://arxiv.org/abs/1202.4893} {arXiv:1202.4893 [gr-qc]} \BibitemShut
  {NoStop}%
\bibitem [{\citenamefont {Kobayashi}\ \emph {et~al.}(2014)\citenamefont
  {Kobayashi}, \citenamefont {Motohashi},\ and\ \citenamefont
  {Suyama}}]{Kobayashi:2014wsa}%
  \BibitemOpen
  \bibfield  {author} {\bibinfo {author} {\bibfnamefont {T.}~\bibnamefont
  {Kobayashi}}, \bibinfo {author} {\bibfnamefont {H.}~\bibnamefont
  {Motohashi}}, \ and\ \bibinfo {author} {\bibfnamefont {T.}~\bibnamefont
  {Suyama}},\ }\href {\doibase 10.1103/PhysRevD.89.084042} {\bibfield
  {journal} {\bibinfo  {journal} {Phys. Rev.}\ }\textbf {\bibinfo {volume}
  {D89}},\ \bibinfo {pages} {084042} (\bibinfo {year} {2014})},\ \Eprint
  {http://arxiv.org/abs/1402.6740} {arXiv:1402.6740 [gr-qc]} \BibitemShut
  {NoStop}%
\bibitem [{\citenamefont {Regge}\ and\ \citenamefont
  {Wheeler}(1957)}]{Regge:1957td}%
  \BibitemOpen
  \bibfield  {author} {\bibinfo {author} {\bibfnamefont {T.}~\bibnamefont
  {Regge}}\ and\ \bibinfo {author} {\bibfnamefont {J.~A.}\ \bibnamefont
  {Wheeler}},\ }\href {\doibase 10.1103/PhysRev.108.1063} {\bibfield  {journal}
  {\bibinfo  {journal} {Phys. Rev.}\ }\textbf {\bibinfo {volume} {108}},\
  \bibinfo {pages} {1063} (\bibinfo {year} {1957})}\BibitemShut {NoStop}%
\bibitem [{\citenamefont {Ganguly}\ \emph {et~al.}(2018)\citenamefont
  {Ganguly}, \citenamefont {Gannouji}, \citenamefont {Gonzalez-Espinoza},\ and\
  \citenamefont {Pizarro-Moya}}]{Ganguly:2017ort}%
  \BibitemOpen
  \bibfield  {author} {\bibinfo {author} {\bibfnamefont {A.}~\bibnamefont
  {Ganguly}}, \bibinfo {author} {\bibfnamefont {R.}~\bibnamefont {Gannouji}},
  \bibinfo {author} {\bibfnamefont {M.}~\bibnamefont {Gonzalez-Espinoza}}, \
  and\ \bibinfo {author} {\bibfnamefont {C.}~\bibnamefont {Pizarro-Moya}},\
  }\href {\doibase 10.1088/1361-6382/aac8a0} {\bibfield  {journal} {\bibinfo
  {journal} {Class. Quant. Grav.}\ }\textbf {\bibinfo {volume} {35}},\ \bibinfo
  {pages} {145008} (\bibinfo {year} {2018})},\ \Eprint
  {http://arxiv.org/abs/1710.07669} {arXiv:1710.07669 [gr-qc]} \BibitemShut
  {NoStop}%
\bibitem [{\citenamefont {De~Felice}\ and\ \citenamefont
  {Tsujikawa}(2012)}]{DeFelice:2011bh}%
  \BibitemOpen
  \bibfield  {author} {\bibinfo {author} {\bibfnamefont {A.}~\bibnamefont
  {De~Felice}}\ and\ \bibinfo {author} {\bibfnamefont {S.}~\bibnamefont
  {Tsujikawa}},\ }\href {\doibase 10.1088/1475-7516/2012/02/007} {\bibfield
  {journal} {\bibinfo  {journal} {JCAP}\ }\textbf {\bibinfo {volume} {1202}},\
  \bibinfo {pages} {007} (\bibinfo {year} {2012})},\ \Eprint
  {http://arxiv.org/abs/1110.3878} {arXiv:1110.3878 [gr-qc]} \BibitemShut
  {NoStop}%
\bibitem [{\citenamefont {Bellini}\ and\ \citenamefont
  {Sawicki}(2014)}]{Bellini:2014fua}%
  \BibitemOpen
  \bibfield  {author} {\bibinfo {author} {\bibfnamefont {E.}~\bibnamefont
  {Bellini}}\ and\ \bibinfo {author} {\bibfnamefont {I.}~\bibnamefont
  {Sawicki}},\ }\href {\doibase 10.1088/1475-7516/2014/07/050} {\bibfield
  {journal} {\bibinfo  {journal} {JCAP}\ }\textbf {\bibinfo {volume} {1407}},\
  \bibinfo {pages} {050} (\bibinfo {year} {2014})},\ \Eprint
  {http://arxiv.org/abs/1404.3713} {arXiv:1404.3713 [astro-ph.CO]} \BibitemShut
  {NoStop}%
\bibitem [{\citenamefont {{Leaver}}(1985)}]{1985RSPSA.402..285L}%
  \BibitemOpen
  \bibfield  {author} {\bibinfo {author} {\bibfnamefont {E.~W.}\ \bibnamefont
  {{Leaver}}},\ }\href {\doibase 10.1098/rspa.1985.0119} {\bibfield  {journal}
  {\bibinfo  {journal} {Proceedings of the Royal Society of London Series A}\
  }\textbf {\bibinfo {volume} {402}},\ \bibinfo {pages} {285} (\bibinfo {year}
  {1985})}\BibitemShut {NoStop}%
\bibitem [{\citenamefont {{Dolan}}\ and\ \citenamefont
  {{Ottewill}}(2009)}]{2009CQGra..26v5003D}%
  \BibitemOpen
  \bibfield  {author} {\bibinfo {author} {\bibfnamefont {S.~R.}\ \bibnamefont
  {{Dolan}}}\ and\ \bibinfo {author} {\bibfnamefont {A.~C.}\ \bibnamefont
  {{Ottewill}}},\ }\href {\doibase 10.1088/0264-9381/26/22/225003} {\bibfield
  {journal} {\bibinfo  {journal} {Classical and Quantum Gravity}\ }\textbf
  {\bibinfo {volume} {26}},\ \bibinfo {eid} {225003} (\bibinfo {year}
  {2009})},\ \Eprint {http://arxiv.org/abs/0908.0329} {arXiv:0908.0329 [gr-qc]}
  \BibitemShut {NoStop}%
\bibitem [{\citenamefont {Hou}\ \emph {et~al.}(2018)\citenamefont {Hou},
  \citenamefont {Gong},\ and\ \citenamefont {Liu}}]{Hou:2017bqj}%
  \BibitemOpen
  \bibfield  {author} {\bibinfo {author} {\bibfnamefont {S.}~\bibnamefont
  {Hou}}, \bibinfo {author} {\bibfnamefont {Y.}~\bibnamefont {Gong}}, \ and\
  \bibinfo {author} {\bibfnamefont {Y.}~\bibnamefont {Liu}},\ }\href {\doibase
  10.1140/epjc/s10052-018-5869-y} {\bibfield  {journal} {\bibinfo  {journal}
  {Eur. Phys. J.}\ }\textbf {\bibinfo {volume} {C78}},\ \bibinfo {pages} {378}
  (\bibinfo {year} {2018})},\ \Eprint {http://arxiv.org/abs/1704.01899}
  {arXiv:1704.01899 [gr-qc]} \BibitemShut {NoStop}%
\bibitem [{\citenamefont {Gong}\ and\ \citenamefont
  {Hou}(2018)}]{Gong:2018ybk}%
  \BibitemOpen
  \bibfield  {author} {\bibinfo {author} {\bibfnamefont {Y.}~\bibnamefont
  {Gong}}\ and\ \bibinfo {author} {\bibfnamefont {S.}~\bibnamefont {Hou}},\
  }\bibfield  {booktitle} {\emph {\bibinfo {booktitle} {{International
  Conference on Quantum Gravity Shenzhen, Guangdong, China, March 26-28,
  2018}}},\ }\href {\doibase 10.3390/universe4080085} {\  (\bibinfo {year}
  {2018}),\ 10.3390/universe4080085},\ \bibinfo {note}
  {[Universe4,no.8,85(2018)]},\ \Eprint {http://arxiv.org/abs/1806.04027}
  {arXiv:1806.04027 [gr-qc]} \BibitemShut {NoStop}%
\bibitem [{\citenamefont {Martel}\ and\ \citenamefont
  {Poisson}(2005)}]{Martel:2005ir}%
  \BibitemOpen
  \bibfield  {author} {\bibinfo {author} {\bibfnamefont {K.}~\bibnamefont
  {Martel}}\ and\ \bibinfo {author} {\bibfnamefont {E.}~\bibnamefont
  {Poisson}},\ }\href {\doibase 10.1103/PhysRevD.71.104003} {\bibfield
  {journal} {\bibinfo  {journal} {Phys. Rev.}\ }\textbf {\bibinfo {volume}
  {D71}},\ \bibinfo {pages} {104003} (\bibinfo {year} {2005})},\ \Eprint
  {http://arxiv.org/abs/gr-qc/0502028} {arXiv:gr-qc/0502028 [gr-qc]}
  \BibitemShut {NoStop}%
\bibitem [{\citenamefont {McManus}\ \emph {et~al.}(2019)\citenamefont
  {McManus}, \citenamefont {Berti}, \citenamefont {Macedo}, \citenamefont
  {Kimura}, \citenamefont {Maselli},\ and\ \citenamefont
  {Cardoso}}]{McManus:2019ulj}%
  \BibitemOpen
  \bibfield  {author} {\bibinfo {author} {\bibfnamefont {R.}~\bibnamefont
  {McManus}}, \bibinfo {author} {\bibfnamefont {E.}~\bibnamefont {Berti}},
  \bibinfo {author} {\bibfnamefont {C.~F.~B.}\ \bibnamefont {Macedo}}, \bibinfo
  {author} {\bibfnamefont {M.}~\bibnamefont {Kimura}}, \bibinfo {author}
  {\bibfnamefont {A.}~\bibnamefont {Maselli}}, \ and\ \bibinfo {author}
  {\bibfnamefont {V.}~\bibnamefont {Cardoso}},\ }\href@noop {} {\  (\bibinfo
  {year} {2019})},\ \Eprint {http://arxiv.org/abs/1906.05155} {arXiv:1906.05155
  [gr-qc]} \BibitemShut {NoStop}%
\bibitem [{\citenamefont {Charmousis}\ \emph
  {et~al.}(2019{\natexlab{a}})\citenamefont {Charmousis}, \citenamefont
  {Crisostomi}, \citenamefont {Langlois},\ and\ \citenamefont
  {Noui}}]{Charmousis:2019fre}%
  \BibitemOpen
  \bibfield  {author} {\bibinfo {author} {\bibfnamefont {C.}~\bibnamefont
  {Charmousis}}, \bibinfo {author} {\bibfnamefont {M.}~\bibnamefont
  {Crisostomi}}, \bibinfo {author} {\bibfnamefont {D.}~\bibnamefont
  {Langlois}}, \ and\ \bibinfo {author} {\bibfnamefont {K.}~\bibnamefont
  {Noui}},\ }\href@noop {} {\  (\bibinfo {year} {2019}{\natexlab{a}})},\
  \Eprint {http://arxiv.org/abs/1907.02924} {arXiv:1907.02924 [gr-qc]}
  \BibitemShut {NoStop}%
\bibitem [{\citenamefont {Charmousis}\ \emph
  {et~al.}(2019{\natexlab{b}})\citenamefont {Charmousis}, \citenamefont
  {Crisostomi}, \citenamefont {Gregory},\ and\ \citenamefont
  {Stergioulas}}]{Charmousis:2019vnf}%
  \BibitemOpen
  \bibfield  {author} {\bibinfo {author} {\bibfnamefont {C.}~\bibnamefont
  {Charmousis}}, \bibinfo {author} {\bibfnamefont {M.}~\bibnamefont
  {Crisostomi}}, \bibinfo {author} {\bibfnamefont {R.}~\bibnamefont {Gregory}},
  \ and\ \bibinfo {author} {\bibfnamefont {N.}~\bibnamefont {Stergioulas}},\
  }\href {\doibase 10.1103/PhysRevD.100.084020} {\bibfield  {journal} {\bibinfo
   {journal} {Phys. Rev.}\ }\textbf {\bibinfo {volume} {D100}},\ \bibinfo
  {pages} {084020} (\bibinfo {year} {2019}{\natexlab{b}})},\ \Eprint
  {http://arxiv.org/abs/1903.05519} {arXiv:1903.05519 [hep-th]} \BibitemShut
  {NoStop}%
\bibitem [{\citenamefont {Suvorov}(2019)}]{Suvorov:2019qow}%
  \BibitemOpen
  \bibfield  {author} {\bibinfo {author} {\bibfnamefont {A.~G.}\ \bibnamefont
  {Suvorov}},\ }\href {\doibase 10.1103/PhysRevD.99.124026} {\bibfield
  {journal} {\bibinfo  {journal} {Phys. Rev.}\ }\textbf {\bibinfo {volume}
  {D99}},\ \bibinfo {pages} {124026} (\bibinfo {year} {2019})},\ \Eprint
  {http://arxiv.org/abs/1905.02021} {arXiv:1905.02021 [gr-qc]} \BibitemShut
  {NoStop}%
\bibitem [{oxw()}]{oxwebsite}%
  \BibitemOpen
  \href {https://www2.physics.ox.ac.uk/contacts/people/tattersall} {\emph
  {\bibinfo {title}
  {https://www2.physics.ox.ac.uk/contacts/people/tattersall}}}\BibitemShut
  {NoStop}%
\end{thebibliography}%

\end{document}